%% file: ms.tex
\documentstyle[12pt,aasms4]{article}  

\received{8 October 1998}
\accepted{23 April 1999}
\lefthead{Frisch et al.}
\righthead{Dust in Local Interstellar Wind}
 
\begin{document}
\title{Dust in the Local Interstellar Wind}
\begin{center}
\today
\end{center}
\author{Priscilla C. Frisch}
\affil{Department of Astronomy and Astrophysics, University of Chicago, Chicago, IL  60637}
\author{Johann M. Dorschner}
\affil{Astrophysikalisches Institut und Universitaets-Sternwarte,
Schillergaesschen 3, D-07745 Jena, Germany}

\author{Johannes Geiss}
\affil{International Space Science Institut, Bern, Switzerland}
\author{J. Mayo Greenberg}
\affil{Leiden Observatory Laboratory, Postbus 9504, 2300 RA
Leiden, The Netherlands}
\author{Eberhard Gr\"un and Markus Landgraf}
\affil{Max-Planck Institut fuer Kernphysik, Heidelberg, Germany}
\author{Peter Hoppe}
\affil{Max-Planck-Institut for Chemistry,
Cosmochemistry Department,
P.O. Box 3060, D-55020 Mainz, Germany}
\author{Anthony P. Jones}
\affil{Institut d'Astrophysique Spatiale,
       Universit\'e Paris XI, B\^atiment 121,
       91405 Orsay Cedex, France}
\author{Wolfgang Kr\"atschmer}
\affil{Max-Planck Institut fuer Kernphysik, Heidelberg, Germany}
\author{Timur J. Linde}
\affil{University of Michigan, Aerospace Engineering, Ann Arbor, MI  48109}
\altaffiltext{1}{Currently at the Department of Astronomy and Astrophysics, University of Chicago, Chicago, IL  60637}
\author{Gregor E. Morfill}
\affil{Max-Planck-Institut fuer extraterrestrische Physik, 85740 Garching, Germany}
\author{William T. Reach}
\affil{Infrared Processing and Analysis Center, California Institute of
Technology, Mail Stop 100-22, Pasadena, CA 91125}

\author{Jonathan D. Slavin\altaffilmark{1,2}}
\affil{Eureka Scientific Inc., 2452 Delmer St. Suite 100, Oakland, CA
94602-3017}
\altaffiltext{1}{Also: University of California, Space Sciences
Laboratory, Berkeley, CA 94720}
\altaffiltext{2}{Postal address: NASA/Ames Research Center, MS 245-3, 
Moffett Field, CA 94035-1000}

\author{Jiri Svestka}
\affil{Prague Observatory, Prague, 11846 Czech Republic}
\author{Adolf N. Witt}
\affil{Ritter Astrophysical Research Center,
University of Toledo, Toledo, OH 43606}
\and
\author{Gary P. Zank}
\affil{Bartol Research Institute, University of Delaware, Newark, DE 19716}

\begin{abstract}

The gas-to-dust mass ratios found for interstellar dust within the
Solar System, versus values determined astronomically for the
cloud around the Solar System, suggest that 
large and small interstellar grains have separate histories, and that large 
interstellar grains preferentially detected by spacecraft are not
formed exclusively by mass exchange with nearby interstellar gas.
Observations by the Ulysses and Galileo satellites of the mass spectrum
and flux rate of interstellar dust within the heliosphere are combined with information about the density, composition, and relative flow speed and
direction of interstellar gas in the cloud surrounding the solar
system to derive an {\it in situ} value for the gas-to-dust mass ratio, 
$R_{\rm g/d}$=94$^{+46}_{-38}$.  This ratio
is dominated by the larger near-micron sized grains.
Including an estimate for the mass of smaller grains, which do not 
penetrate the heliosphere due to charged grain interactions with
heliosheath and solar wind plasmas, and including estimates for the
mass of the larger population of interstellar micrometeorites,
the total gas-to-dust mass ratio in the cloud surrounding the Solar System 
is half this value.
Based on {\it in situ} data, interstellar dust grains in the 
of 10$^{-14}$ to 10$^{-13}$ g mass range are underabundant in the 
Solar System, compared to an MRN mass distribution scaled to the local
interstellar gas density, because such small grains do not penetrate the 
heliosphere.
The gas-to-dust mass ratios are also derived by combining spectroscopic 
observations of the gas-phase abundances in the nearest interstellar clouds.
Measurements of interstellar absorption lines formed in the cloud around the solar system, 
as seen in the direction of $\epsilon$ CMa,
give $R_{\rm g/d}$=427$^{+72}_{-207}$ for assumed solar reference abundances,
and $R_{\rm g/d}$=551$^{+61}_{-251}$ for assumed B-star reference abundances.
These values exceed the {\it in situ} value, suggesting either
grain mixing or grain histories are not correctly understood, or
that sweptup stardust is present.
Such high values for diffuse interstellar clouds 
are strongly supported by diffuse cloud data seen towards $\lambda$ Sco
and 23 Ori, provided B-star reference abundances apply.
If solar reference abundances prevail, however, the surrounding cloud 
is seen to have greater than normal dust destruction compared to 
higher column density diffuse clouds.
The cloud surrounding the Solar System exhibits enhanced gas-phase
abundances of refractory elements such as Fe$^+$ and Mg$^+$, indicating
the destruction of dust grains by shock fronts.
The good correlation locally between Fe$^+$ and Mg$^+$ indicates that the gas-phase
abundances of these elements are dominated by grain destruction,
while the poor correlation between Fe$^+$ and H$^{\rm \circ}$ indicates either
variable gas ionization or the decoupling of neutral gas and
dust over parsec scalelengths.
These abundances, combined with grain destruction models, indicate
that the nearest interstellar material has been shocked with shocks of
velocity $\sim$150 km s$^{-1}$. 
If solar reference abundances are correct, the low $R_{\rm g/d}$
value towards $\lambda$ Sco may indicate that at least one cloud
component in this direction contains dust grains which have retained their
silicate mantles, and are responsible for the polarization of the light
from nearby stars seen in this general region.
Weak frictional coupling between gas and dust in nearby low density gas permit
inhomogeneities to be present.  

\end{abstract}


\section{Introduction}\label{intro}
The direct detection of interstellar dust grains within the solar
system by the Ulysses and Galileo satellites provides an opportunity
to constrain the properties, history, and origin of these grains
by comparing {\it in situ} and remote data.  
The {\it in situ} satellite data sample the large end of the grain size distribution, 0.1
$\leq$ $a_{\rm gr}$ $\leq$ $\sim$4 $\mu$m ( $a_{\rm gr}$ is grain radius),
since small grains are excluded from the
inner heliosphere by interaction with the solar wind and in the
heliopause region 
(\cite{levy76,wallis,mann}, \S\ref{spacecraft} ).  
In contrast, optical and ultraviolet observations of extinction
and polarization of starlight by interstellar dust grains give
information on dust grains in size ranges 0.001$<$$a_{\rm gr}$$<$0.3
$\mu$m (e.g. Mathis 1990).
The goal of this paper is to compare the properties of the interstellar
dust grains inferred from {\it in situ} data with
grain properties derived from observations of interstellar matter (ISM)
in front of nearby stars, in order to understand
the properties of interstellar dust grains in nearby interstellar gas 
and in the cloud surrounding the Solar System.  
Comparisons with presolar grains in meteorites, and interplanetary dust grains,
highlight similarities and differences compared to interstellar grains.  
These combined perspectives provide an interdisciplinary glimpse of the
dust grains embedded in the interstellar cloud flowing through the Solar System, or the ``Local Interstellar Wind'' as it has been nicknamed.
In this paper, we will denote the interstellar cloud surrounding the
Solar System as the ``Local Interstellar Cloud'' (LIC).  The
downstream (or, equivalently, downwind) directions are given in Appendix \ref{ap_LIC}.

An interdisciplinary approach for understanding interstellar dust grains is
employed, based on comparing 
{\it in situ} spacecraft data (\S\ref{spacecraft})
and astronomical data (\S\ref{LICdust}) to evaluate the gas-to-dust
mass ratio in the LIC.
The criteria for separating interstellar dust grains from interplanetary dust grains observed
by the Ulysses and Galileo satellites are reviewed (\S\ref{identifications}),
The impact events selected from the combined 
Galileo and Ulysses interstellar dust datasets
(\S\ref{datasets})
provide a data set from which the mass distribution of the {\it in situ} 
detected events can be determined and compared with
the predictions of standard astronomical models (\S\ref{massdis}). 
Such a comparison is original to this paper.
These events then provide a basis for estimating the gas-to-dust
mass ratio for detected grains (\S\ref{totaldens}).  
Alternative methods of obtaining information on interstellar dust grains in the
Solar System are discussed, including possible thermal infrared
emission features and excess carbon from pickup ion observations (\S\ref{alternative}).
Gas-dynamic models are used to evaluate the exclusion of small interstellar grains at the heliopause and by interactions with the solar wind within the heliosphere (\S\ref{exclusion}),
as a function of the adopted
heliosphere model (\S\ref{mhdmodel}) and grain charge and gyroradii
(\S\ref{graincharge}, \ref{hpfiltration}).
In Section \ref{transition} it is shown that grain destruction processes
are unlikely to be significant in the time it takes for grains to traverse
the region between a heliospheric bow-shock and the
heliopause.  The dynamics of interstellar dust grains within the heliosphere, including
the counter-effects of radiation pressure and gravitational
focusing, are discussed in Section \ref{hsfiltration}.
Section \ref{LICdust} investigates the properties of interstellar dust grains in interstellar material
near the Sun, including grain properties derived from mass ``missing'' from the gas phase (\S\ref{licgas2dust})
when absorption line data are compared with the nominal reference
abundance for the material. 
Astronomical observations of $\epsilon$ CMa and $\lambda$ Sco, combined with an
assumed reference abundance for the gas, give a range for the values of the
gas-to-dust ratio in the LISM which constrain either the
reference abundances or the gas-dust mixing history (\S\ref{licgas2dust}).  
Enhanced abundances of refractories (\S\ref{licgas2dust}) and inhomogeneous abundances in the LISM are present in the LISM (\S\ref{licgas2dust},\ref{lism_abs}).
These are consistent with the destruction of dust grains by interstellar
shocks, and grain destruction models yield the shock velocity (\S\ref{SF}).
Presolar grains in meteorites may constitute only a small fraction of the interstellar particles, 
but are discussed in search of insights about the grain size
and distribution and composition (\S\ref{hoppe}).  
In Section \ref{disc} we discuss the implications of our comparison
between {\it in situ} versus remotely detected interstellar dust grains in the LISM.  
We will conclude that dust grains are overabundant in the cloud feeding interstellar grains into the
Solar System, in comparison to the sight-line averaged LIC cloud
component observed towards $\epsilon$ CMa.
The higher column density diffuse clouds towards
$\lambda$ Sco (and 23 Ori) are consistent with the {\it in situ} 
results, however, indicating possible homogeneities.  Consequently, we look
at the dynamic separation of grains and gas (\S\ref{dynsep}),
and consider the possibility that a portion of the grain population has not 
exchanged material with the gas phase.
Some consequences for dust grain evolution follow from the
detected large mass tail on the interstellar dust grain mass
distribution (\S\ref{evolution}), and grain models (\S\ref{coremodel}).
Finally, we discuss constraints placed on the reference abundance of 
interstellar gas by the comparisons between {\it in situ} and
astronomical data (\S\ref{licgas2dust}).

In several sections of the paper, for ease in reading, supporting
material is given in appendices.
Appendix \ref{ap_LIC} discusses LIC properties.  More detail is given
on the upstream direction that provided the basis for selecting
the {\it in situ} events (Appendix \ref{direction}).  
Observations of the weak polarization of optical light towards nearby stars are the only direct data on nearby interstellar dust grains and are summarized (Appendix~\ref{polarization}).
The interstellar absorption line data, drawn from the literature, which provide the basis for the
understanding the LIC dust component discussed in \S\ref{LICdust} are 
presented in Appendices \ref{ap_depl} and \ref{LICcolumns}.

Unless otherwise indicated,
interstellar dust grains are assumed to be spherical with density $\rho$=2.5 g cm$^{-3}$.
The conclusions in this paper, which focus on differences in the
gas-to-dust mass ratios determined from {\it in situ} versus 
astronomical data, generally are not affected by this choice of the
bulk grain density, which is somewhat uncertain.  The main use of
this bulk density is to provide an estimate of the grain-size range which
corresponds to the measured grains, for a limited grain model.
Exploring other grain models is beyond the scope of this paper.
 
\section{Interstellar Cloud Surrounding the Solar System}\label{LIC_model}
The comparison of dust grains observed inside versus outside of the heliosphere
requires an understanding of both the density of interstellar gas in the
cloud surrounding the Solar System, and of the exclusion of charged interstellar
grains from the heliosphere.  
Gas density is required to calculate the gas-to-dust mass ratio for the {\it in situ} grain population.  
Heliosphere models are constrained by solar 
wind ram pressure and by the density, ionization, temperature and magnetic field of the surrounding cloud.
The interstellar magnetic field strength governs the exclusion of the smallest
particles from the heliosphere.
The physical properties of the diffuse warm low density cloud surrounding the Solar System are
summarized in Table \ref{licprop}; the basis for the selection of these values
is discussed in Appendix \ref{ap_LIC}.  A number of primary 
importance to the results of this paper is the total space density
of the cloud feeding dust grains into the heliosphere, as this is the
number to which the {\it in situ} data are normalized.  
Based on a range of data and models, this number is unlikely to be
more than 50\% larger than the adopted value, and the uncertainty
does not alter the conclusions of this paper.

%\placetable{licprop}
\input{tabLIC.tex}

The cloud surrounding the Solar System is a fragment of a cloud complex
sweeping past the Sun from the direction of the center of Loop I
(\cite{fr95}).  Within about 35 pc of the Sun, column densities
do not exceed $\sim$10$^{19}$ cm$^{-2}$, and are an order of magnitude
larger in the galactic center hemisphere than the anti-center hemisphere.
The material within about 35 pc of the Sun is referred to as 
local interstellar matter (LISM).

\section{Interstellar Dust Properties from Spacecraft {\it In Situ} Observations}\label{spacecraft}
Gr\"un et al. (1994) have shown that the direction of motion of
interstellar dust grains is consistent with the direction of motion 
of interstellar
He$^{\rm o}$ also measured with Ulysses (\cite{witte93}).  
Appendix \ref{direction} provides the criteria by which interstellar dust grains are
selected from the Ulysses/Galileo data sets.
With a larger data sample, we will show (Appendix \ref{direction}) that this
result holds when a quantitative $\chi^2$-fit of the data is
performed.  Gr\"un et al. (1994) determined the mean number-flux of
interstellar grains through the heliosphere as 
$1.5\cdot 10^{-8} {\rm cm}^{-2}\ {\rm s}^{-1}$, 
and that the most abundant grains in these data
have masses of $3\cdot 10^{-13}\ {\rm g}$, corresponding to grain
radius $a_{\rm gr}$$\sim$0.3 $\mu$m for density 2.5 g cm$^{-3}$
and spherical grains.
We will compare the observed grain mass
distribution with the standard MRN model mass distribution for 
interstellar dust,
and show that in the detected population of grains with $a_{\rm gr}$$>$0.3 $\mu$m
contribute most of the mass. 
The exclusion of smaller grains by heliospheric interactions is
discussed in Section \ref{hpfiltration}.

\subsection{Criteria for Identifying Interstellar Dust Grains}\label{identifications}
One of the prime objectives of the Ulysses dust detector was to search
for interstellar dust, which must be distinguished from interplanetary
dust particles (\cite{gruen92a}). The main distinction between these
two dust populations was thought to be their respective spatial
distributions: interplanetary dust is strongly concentrated to the
ecliptic plane while the interstellar dust flux 
should show little dependence on ecliptic latitude.  However, it was
fortuitous that interstellar dust was even more easily
identified. Delays in the launch date of Ulysses lead to an
out-of-ecliptic orbit that was perpendicular to the flow of
interstellar dust.  Thereby, the spin of the spacecraft gives rise to
the maximum modulation of the interstellar impact rate as a function
of spin angle and allows an accurate determination of flow
direction. In addition the aphelion of the Ulysses orbit was on the side of
the Sun where interstellar flow was opposite to the orbital motion of
interplanetary dust. 
Thus, the interplanetary dust flux 
which displays a
strong radial gradient was at minimum and the interstellar flux was the
dominant flux. For a launch date three years earlier as originally
planned, little interstellar dust would have
entered the detector because then the descending node of Ulysses'
orbit would be directed parallel to the upstream direction. In such a
configuration the dust detector points would always point nearly perpendicular to
the interstellar direction.

Micron-sized interstellar dust particles passing through the solar
system have been positively identified and separated from
interplanetary dust by meeting three criteria (\cite{gruen94,baguhl96}):
\begin{enumerate}
\item After the Jupiter flyby Ulysses observed a dominant flux of dust
particles arriving on prograde orbits from the opposite direction of interplanetary
dust.  The detected impact direction is compatible
with grains from outside the Solar System, if we assume that these
grains enter the Solar System from close to the upstream direction of
the interstellar wind (\cite{witte93}).  The determination of the
upstream direction of interstellar dust is discussed in
Appendix~\ref{direction} (see also Baguhl et al. 1995).

\item Despite a big uncertainty in the impact speed determination
(factor 2, \cite{gruen92}) by the Ulysses dust detector, most
particles had speeds in excess of the heliospheric escape speed. To
detect high impact velocities we use the total impact charge measured
by the detector. The total charge produced by an impact is highly
sensitive to the impact velocity (\cite{gruen92}). High impact charges
can also be produced by slow interplanetary grains with large
masses. The flux of these grains is low beyond $3\ {\rm AU}$, so
we can neglect them when considering data that were collected at the
5 AU heliocentric distance of Jupiter.

\item As discussed above, the flux of these interstellar particles was independent of ecliptic
latitude (\cite{landgraf96}) in contrast to interplanetary dust that is
strongly concentrated towards the ecliptic plane and the inner solar
system. Dust emanating from the Jovian system is localized to the
Jovian vicinity (\cite{gruen93}).
\end{enumerate}

Consideration of these factors demonstrates that these {\it in situ}
particles are not interplanetary dust particles.
What about other more exotic Solar System dust populations, 
like dust from the Kuiper belt or the Oort cloud?

The Kuiper belt, thought to be the source of the short period comets, is
located outside the orbit of Neptune (\cite{flynn94,jewitt93,jewitt95}). It is
considered to be the 
remnant of the outer parts of the protoplanetary disk, and
consequently the Kuiper belt objects are concentrated near the
ecliptic plane and move in nearly circular orbits with typical radii
of $40-80\ {\rm AU}$. Thus, most of the dust
produced in the Kuiper belt, if it should reach the inner heliosphere
at all, would come from directions near the ecliptic plane and from all heliospheric longitudes, which is
not compatible with the dust observations at medium and high solar
latitude. The same argument can be used to exclude other potential
Solar System sources located inside the heliosphere.

Another population that could mimic the kinematic parameters of
interstellar dust are grains originating in the Oort cloud of comets.  A very
straightforward argument negates any likelihood of the Oort cloud as a
significant source of dust particles in the inner heliosphere. The
Oort cloud occupies a spherical shell at an average solar distance of
$R_{\rm Oort} \approx 5\cdot 10^4\ {\rm AU}$. With an estimated total
mass ($M_{\rm Oort}$) of 20--40 Earth masses (cf. Weissman 1996)
it is considered to be the
source of the long-period comets. Mutual collisions and perhaps other
mechanisms operating in this cloud are producing small grains that
could become accommodated in the interstellar gas flow and reach the
inner heliosphere from the same direction as the gas. However, the
contamination of grains from the Oort cloud in the interstellar grain
flux must be very small, as the following argument shows. If the total
mass flux of interstellar dust of $f_{m,{\rm IS}} = 2\cdot 10^{-20}\
{\rm g}\ {\rm cm}^{-2}\ {\rm s}^{-1}$ (see section
\ref{totaldens}) should all come from the Oort cloud, its life-time
would be given by
\begin{equation}
     t_{\rm Oort}=\frac{M_{\rm Oort}}
	{4\pi R_{\rm Oort}^2 f_{m,{\rm IS}}}.
\end{equation}

The result is $t_{\rm Oort} < 10^5\ {\rm years}$, showing that
the Oort cloud would survive only a very short time if a significant
fraction of the observed interstellar flux were produced by it.

\subsection{Characteristics of Interstellar Dust in the Heliosphere}

\subsubsection{The Galileo and Ulysses Interstellar Dust Datasets}\label{datasets}
Using the criteria for interstellar impacts given in section 3.1, we
define the subsets of the Galileo and Ulysses data which make up the
interstellar dust datasets. For this, we use Ulysses data after Jupiter flyby
until March 1996, excluding the ecliptic crossing, and Galileo data on
its way to Jupiter between $3.5\ {\rm AU}$ and the arrival in the
Jovian system. The dataset was collected either
in-ecliptic outside $3.5\ {\rm AU}$ or at higher heliocentric
latitudes. In both the Galileo and Ulysses data, impacts caused by
Jupiter stream particles have been identified and removed
Gr\"un et al. (1993;1996b).

The dataset from which the interstellar dust grain mass distribution
will be derived (\S\ref{massdis}) was selected by grain impact direction. The criterion
is: All impacts that were not identified as Jupiter stream particles
(\cite{gruen93}), and that are compatible with coming from within $\pm 70^\circ$
of the upstream direction of the interstellar wind (ignoring possible
deflections by heliospheric interaction) plus a $10^\circ$-margin for
uncertainties in the upstream direction,
are candidates for interstellar impacts. Since Ulysses returned to the
inner Solar System, $\beta$-meteoroids and interplanetary grains have
to be removed from the data collected by Ulysses in this phase of its
mission within a few AU of the Sun. Possible impacts
from $\beta$-meteoroids originating above the poles were
removed by restricting the impact charge to be 
larger then $1\cdot 10^{-13}\ {\rm C}$. In the
Ulysses dataset, impacts by interplanetary grains on orbits with small
inclinations (as reported by Gr\"un et al. 1997) are removed by
cutting out the data that were collected 
$\pm 60^\circ$ around the ecliptic crossing, because in this phase of the
mission Ulysses was on the Sun side where interstellar dust and interplanetary dust grains
came from the same direction. 
The remaining set of detected impacts defines our interstellar dataset for the
determination of the mass distribution. This dataset contains $337$
events detected by Galileo. We note that for $28$ of them, the impact direction was
not detected. Ulysses contributes $313$ particles, $8$ of them without
direction information.  For the determination of the upstream
direction we have to choose the impact velocity criteria so as not to
introduce a directional bias.  The impact velocity criterion for the
Ulysses dataset
from which the upstream direction is derived is that the event, is not
identified as a Jupiter stream event and that the impact charge has to
be larger than $1\cdot 10^{-13}\ {\rm C}$. This means that if the
impacting particle had a mass of $10^{-13}\ {\rm g}$, the impact
velocity has to be larger than $20\ {\rm km}\ {\rm s}^{-1}$. 
In order to have a
constant relationship between measured rotation angle and upstream
direction we chose the period of time between Jupiter flyby in
February 1992 and May 1993 where the geometry didn't change
significantly. During this phase Ulysses was outside $3\ {\rm AU }$.

\subsubsection{Mass Distribution }\label{massdis}
\begin{figure}[ht]
\epsscale{0.8}
\plottwo{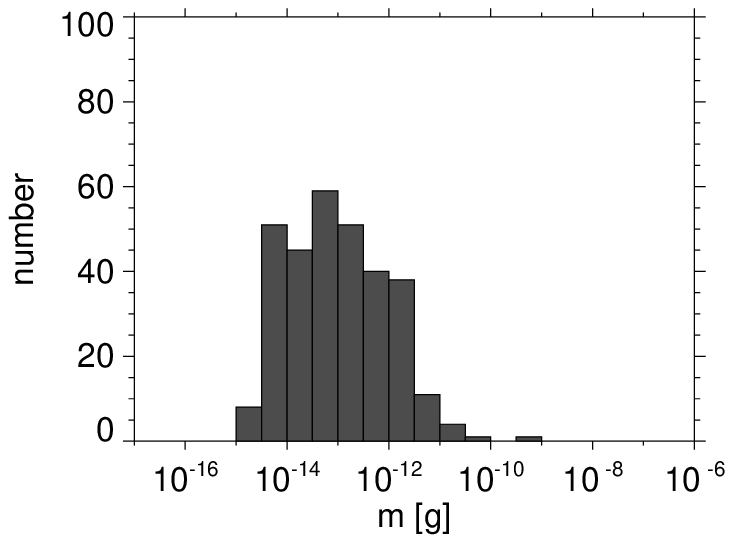}{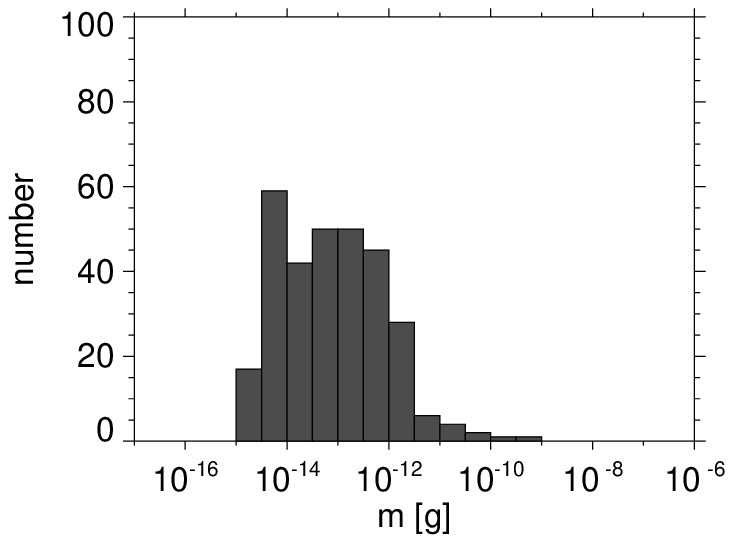}
\caption{\label{massdis_fig}\it Mass histograms of interstellar grains
detected by (a) Galileo with $309$ impacts, and (b) Ulysses with $305$
impacts.  These histograms plot the number of interstellar grains per logarithmic
mass interval.  The drop in count rates at masses below $10^{-15}$ g 
(corresponding to grain radius 0.07 $\mu$m for spherical grains with 
density 2.5 g cm$^{-3}$) results from a decrease in
instrument sensitivity for 20 km s$^{-1}$ grains.}
\end{figure}

Figures \ref{massdis_fig} (a) and (b) show the number of interstellar
grains per logarithmic mass interval
detected by Galileo and Ulysses.
%\placefigure{massdis_fig}

We determine the mass of the grain that caused the detected event by
using an impact speed of $26\ {\rm km}\ {\rm s}^{-1}$ and calculate the
mass from the impact charge amplitude using the calibration given in
Gru\"n et al. (1995).
The detection mass threshold for grains with impact velocities of more 
than $20\ {\rm km}\ {\rm s}^{-1}$ is $10^{-15}\ {\rm g}$.
The mean masses are calculated as sample means, i.e. 
the total mass of the detected grain sample
divided by the number of events, 
they are: $\overline{m} = 2\cdot 10^{-12}\ {\rm g}$ for Galileo and $\overline{m} = 1\cdot 10^{-12}\ {\rm g}$ for Ulysses data 
(for $a_{\rm gr}$$\sim$0.5--0.6 $\mu$m for $\rho$=2.5 g cm$^{-3}$ and
spherical grains). The mean masses are larger than the typical
masses reported by Gr\"un et al (1994). 
($m_{\rm typ} = 3\cdot 10^{-13}\ {\rm g}$), because the heavy grains dominate the
total mass.  
The two histograms do not differ significantly, so we combine
both datasets to calculate the mass-density distribution. Small
particles (one or two orders of magnitude above the detection
threshold) are under-abundant compared to a MRN distribution (see below) even if we do not introduce a minimum
mass for the identification of interstellar particles (c.f. Gr\"un et
al. 1994). This is the result of filtration of small
grains by the solar wind magnetic field 
(\S\ref{hsfiltration}) and in the heliopause region (\S\ref{hpfiltration}). 
As expected, the number of particles drops steeply for large masses 
(Figure \ref{massdis_fig}).

The observed grain population is compared with
the expected interstellar dust grain population, using
the empirically derived MRN size distribution (\cite{mathis77}) as a strawman example.
For comparison, the size distribution of Kim, Martin and Hendry (\cite{kim94})
looks like the MRN distribution for smaller grain sizes, turns 
over at radii of $\sim$0.2 $\mu$m, and falls steeply out to 1 $\mu$m radius.
The MRN size distribution is given by
\begin{equation}
\frac{dn}{da_{\rm gr}}=n_{\rm H}{\cal N} a_{\rm gr}^{-\alpha}\label{eq-mrn},
\end{equation}
with $\alpha=3.5$, ${\cal N}$ is a normalization constant, $n_{\rm H}$
the number-density of hydrogen atoms, and the grain radius $a_{\rm gr}$ ranges from
$5\ {\rm nm}$ to $250\ {\rm nm}$. We use $n_{\rm H} = 0.3\ {\rm
cm}^{-3}$ (cf. \S\ref{totaldens}) and ${\cal N}$ is chosen to
fix the MRN gas-to-dust mass ratio to $100$. To compare the {\it in situ}
distribution with the MRN distribution we consider the differential
distribution of mass-density per logarithmic mass interval $n_m(m)$.
For grains with a MRN size distribution this mass distribution
(mass-density per logarithmic mass interval) is given by
\begin{eqnarray}
n_{\rm MRN}(m) & = & \frac{mdn}{d(\log m)} = \frac{\ln 10 {\cal N}
n_H} {3} \left( \frac{3} {4 \pi \rho_{\rm gr}} \right)^{\frac{1 -
\alpha} {3}} m^{\frac{4 - \alpha} {3}}\label{nm_MRN}.
\end{eqnarray}
We assume spherical grains with bulk density $\rho_{\rm gr}=2.5\ {\rm g}\ {\rm cm}^{-3}$.

From equation (\ref{nm_MRN}) follows that for $\alpha = 4$ the mass-density
per logarithmic mass interval is constant.

%\begin{figure
%\plotone{densdis.ps}
%\caption{}
%\end{figure}

\begin{figure}[ht]
\plotone{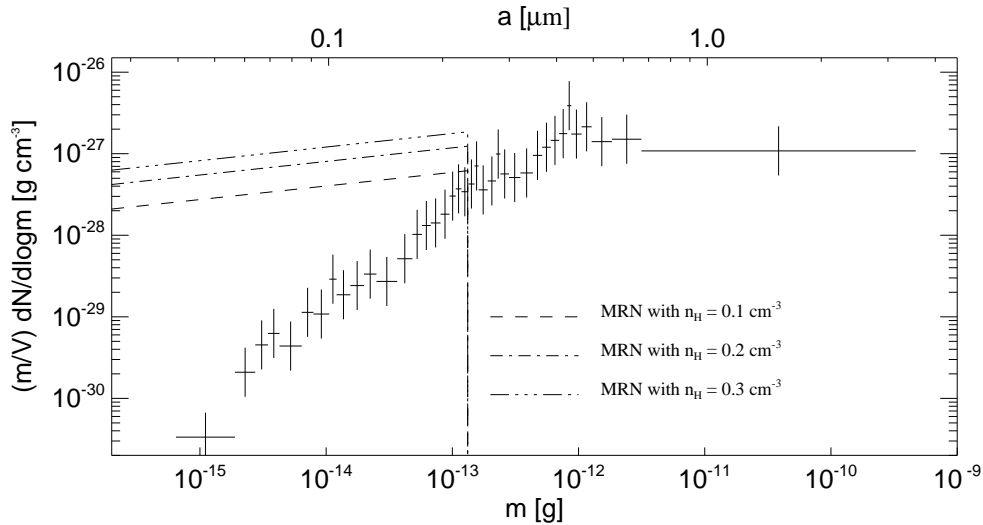}
\caption{\it \label{densdis} Mass-density distribution per logarithmic
mass interval of {\em in situ} particles. The dashed lines show the
MRN distribution for three normalizations of the total mass density in
the LIC, assuming a gas-to-dust mass ratio of $100$. The vertical line
gives the upper limit cutoff of the MRN distribution at 0.25 $\mu$m,
for spherical grains and assumed density 2.5 g cm$^{-2}$.  The crosses
give the value of the distribution function with error-bars which
indicates the uncertainty of mass determination. Each data-point
represents a collection of $16$ impact events.  The top axis gives the
grain radius corresponding to the mass shown on the bottom axis, for
spherical grains.}
\end{figure}

The mass distribution of all {\it in situ} detections shown in Figure
\ref{densdis} does not indicate any steep cutoff for particles with
masses larger than $1.6\cdot 10^{-13}\ {\rm g}$ (or $a_{\rm gr}=0.25 ~ \mu$m for
spherical grains with $\rho_{\rm gr} = 2.5$ g cm$^{-3}$) as postulated by the MRN
distribution. There is limited overlap of the MRN and the {\it in situ}
distributions in the interval $10^{-15} {\rm g}
\leq m_{\rm gr} \leq 1.6\cdot 10^{-13}\ {\rm g}$, but in general the 
number of observed grains in this 
interval is much lower than predicted by the MRN
distribution. For $m_{\rm gr} \geq 1.6\cdot 10^{-13}\ {\rm g}$ the {\it in situ}
distribution seems to extrapolate a MRN like power-law to grains with
larger masses. For masses larger than $10^{-12}\ {\rm g}$ the {\it in situ}
distribution flattens and shows no further increase. At the low-mass end
of the distribution in Figure \ref{densdis}, the events shown are above the sensitivity
cutoff of $\sim$10$^{-15}$ g.  Since the total
mass-density is given by the integral over the distribution function
in Figure \ref{densdis}, it is clear that the large particles 
measured {\it in situ} dominate the total mass, as if also the 
case for an MRN size distribution of particles.
The anomalies at both the small and large ends of the observed grain
population, when compared to predicted sizes based on starlight reddening
measurements, leads us to examine later the screening of
small charged grains by heliospheric plasmas (\S\ref{exclusion}) and
the decoupling of large grains from interstellar grains (\S\ref{SF}, \S\ref{dynsep}).

%\placefigure{densdis}

\subsection{LIC Gas-to-Dust Mass Ratio Implied by {\it In Situ}
Measurements}\label{totaldens}
In this section we derive the gas-to-dust mass ratio implied by the
{\it in situ} detections.  We also estimate the total mass of dust particles
with masses greater than the mass of the heaviest observed grains in the Ulysses
and Galileo datasets, $\sim$10$^{-9}$ g, making a guess about the mass
distribution at the large end.
The exclusion of small grains from the heliosphere by
Lorentz force interactions with the solar wind and in the heliopause region
is discussed (\S\ref{exclusion}).

First we calculate the total mass-density $N_{m,{\rm UG}}$ of interstellar
grains measured by Ulysses and Galileo by integrating the mass
distribution function shown in Figure \ref{densdis}. It follows that
$N_{m,{\rm UG}} = (7.5 \pm 0.8) \cdot 10^{-27}\ {\rm g}\ {\rm cm}^{-3}$, 
which translates to a gas-to-dust mass ratio 
$R_{{\rm g}/{\rm d}}= (n_{\rm H} m_{\rm H} + n_{\rm He} m_{\rm He})/N_{m,{\rm UG}}$ 
of $R_{{\rm g}/{\rm d}} = (94^{+46}_{-38})$ 
assuming $n_H = (0.3\pm0.1)\ {\rm cm}^{-3}$, where $n_{\rm H}$ and $n_{\rm He}$
are the number
densities (cm$^{-3}$) of H and He respectively, $m_{\rm H}$, $m_{\rm He}$
are the respective masses of H and He, and $n_{\rm H}$/$n_{\rm He}$=0.1.  
Note that this determination of the gas-to-dust ratio is independent of any assumptions about the reference abundance of the interstellar gas except
that $n_{\rm H}/n_{\rm He}$=10, 
but it does assume that all helium is in the gas-phase.
If we assume a relative velocity of the
interstellar dust of $26\ {\rm km}\ {\rm s}^{-1}$
(e. g. Witte et al. 1996), we get a dust mass flux of  
$(2.0 \pm 0.2) \cdot 10^{-20}\ {\rm g}\ {\rm cm}^{-2}\ {\rm s}^{-1}$.
The $R_{{\rm g}/{\rm d}}=94$ value compares favorably with generic
gas-to-dust ratios in the interstellar medium (e.g. \cite{spitzer78}).

The extrapolation of the total mass-flux of interstellar grains observed
within the Solar System to the
densities in the interstellar cloud feeding this material into the Solar System
leads to a lower limit, because if we assume that
grains with masses smaller than $3\cdot 10^{-14}\ {\rm g}$ are missing
from the {\em in situ} data because of Lorentz force filtration
in the heliosheath and heliosphere (see section 3.2.3), the mass of these
small grains must be added back to obtain the total mass of dust in the
interstellar medium. 
Furthermore we note that focusing or defocusing effects
by interaction with the solar wind magnetic field as described in
section \ref{hsfiltration} do not affect the large 
($a_{\rm gr} > 1\ {\rm \mu m}$) grains which dominate the total mass-density. We also
do not need to take into account gravitational focusing, because
during the phases when the Galileo and Ulysses data were collected,
both spacecraft were far away from the region in which gravitational
focusing can occur. 

We now evaluate the total mass in the large end of the mass distribution.
To do so, we correct for the mass of the excluded grains, with the
hypothesis that the excluded grains obey a MRN mass distribution.
We assume that the real mass distribution of dust grains in the LIC follows the
MRN law for grains with $m_{\rm gr} \leq 1.6\cdot 10^{-13}\ {\rm g}$, and for
heavier grains that the mass distribution function approaches a constant
level of $1.5\cdot 10^{-27}\ {\rm g}\ {\rm cm}^{-3}$ per logarithmic
mass interval. The total mass-density of grains in the LIC will be the sum of 
$5\cdot 10^{-27}\ {\rm g}\ {\rm cm}^{-3}$ from the MRN-sized grains 
(a number which includes the mass of the excluded grains)
and $1.5\cdot 10^{-27}\ {\rm g}\ {\rm cm}^{-3}$ from every mass decade to which the mass
distribution is extrapolated. For example if we extrapolate to the
radar meteoroids discovered by Taylor et al. (1996)
with $m\approx 10^{-6}\ {\rm g}$, the contribution of
the large grains would be $1\cdot 10^{-26}\ {\rm g}\ {\rm cm}^{-3}$,
which is comparable to the mass-density of the particles detected
{\it in situ}.  The total cross section density of the extended size distribution
increases only a few per cent over that of the MRN distribution 
($1.8\cdot 10^{-22}\ {\rm cm}^2\ {\rm cm}^{-3}$). Therefore
the extinction, emission or scattering characteristics of interstellar
dust are not much changed by the existence of the big particles.
The maximum entropy extinction fitting procedure of Kim and Martin (1996)
for dense clouds seems to require the presence of particles as large as
10 $\mu$m.

In order to estimate the total missing mass due to heliosheath
and heliospheric filtration of charged grains, we look at the difference 
between the MRN curve and observed mass distribution over the 
entire measured mass range.
This is equivalent to calculating the area between the upper predicted MRN-line in Fig.
\ref{densdis} and the observed particles (crosses in Fig. \ref{densdis}) of the distribution of the {\it
in situ} particles between masses of 10$^{-15}$ g and 1.6$\cdot10^{-13}$ g.
The result is that $98\%$ of the number of the assumed MRN particles
(equivalent to $4.9 \cdot 10^{-27}\ {\rm g}\ {\rm cm}^{-3}$) are
removed by interactions of grains with
the solar wind magnetic field and heliosheath plasma. These numbers 
are sensitive to the assumed input initial mass distribution for 
dust grains in the LIC, which is poorly known theoretically.
 
The smaller particles that are excluded from entering the
heliosphere at the transition region ($a_{\rm gr}<0.01$ $\mu$m, see \S\ref{hpfiltration})
also have masses smaller than the $10^{-15}\ {\rm g}$ detection threshold of the Ulysses and Galileo
dust detectors (\cite{gruen92}).

\subsection{Alternative Methods to get Information on Interstellar Dust in the
Solar System}\label{alternative}
\subsubsection{Thermal Emission}

Since the grains detected by Ulysses and Galileo are heated by the Sun,
in principle the infrared radiation they emit should be observable
(e. g. \cite{grogan96}). The
emission depends on the grain temperature and composition. Assuming
the emissivity is independent of wavelength, the equilibrium temperature
would be $278/\sqrt{R_{\rm sun}}$ where $R_{\rm sun}$ is the distance from the Sun in AU.
Taking into account the optical properties of astronomical silicates as
given by Draine and Lee (1984), the temperature is enhanced by up to
50\% for grains in the mass range $10^{-15}\leq m \leq 10^{-9}$ g as
detected by the spaceprobes. 

The emission spectra of graphite and silicate
interstellar grains was calculated by integrating over the line of sight,
taking into account the temperature variation with distance from the Sun
and assuming a uniform spatial distribution. The spectrum of graphite
grains is rather featureless, while the silicate spectrum shows
prominent 10 $\mu$m and 20 $\mu$m features
which come from the hot grains near the Sun. 
The emission of grains in the outer Solar System where the {\it in situ} 
measurements have been made, dominates the spectrum at about $100\ {\rm \mu m}$.

The zodiacal light is $1000$ times greater than the emission we predict
here for interstellar dust heated by the Sun. There remain two
observational methods to detect this emission, however. First, the
thermal emission from zodiacal cloud particles can be significantly decreased by 2 orders of magnitude
by observing from the outer Solar System regions at distances greater than
$3\ {\rm AU}$ from the Sun. Several space telescopes
proposed both to NASA and the ESA would observe from this vantage
point, where the emission of interstellar dust heated by the Sun would
no longer be negligible.  The second method is to search for distinct
signatures of the dust. If the interstellar particles are composed of
silicate material, then they are predicted to have strong 10 and 20
micron emission bands. The $20\ {\rm \mu m}$ band is so wide and close
to the peak of the continuum spectrum that it probably could not be
distinguished, but the $10\ {\rm \mu m}$ feature is more promising. 
Interplanetary grains have a 10 $\mu$m silicate emission feature that
is weaker than interstellar grains (\cite{reach96}).
However, interplanetary silicates may also contribute to these
features, making the identification of emission from interstellar 
grains difficult.
In addition to the spectral signature, there are possible spatial and
temporal signatures due to the interaction of the solar wind and the
particles (\cite{grogan96}) and gravitational focusing.

\subsubsection{Pickup Ion Source}

An opportunity to study the composition of the interstellar grains in
the Solar System is possibly provided by a phenomenon, that the grains
evaporate close to the Sun and the atoms get picked up by the solar
wind after they have been ionized.  Geiss et al. (1996) and Gloeckler and Geiss (1998)
reported the detection of singly ionized carbon and nitrogen pickup ions by the
Ulysses/SWICS experiment coming from an inner source 
near the Sun. Gloeckler and Geiss (1998) noted that four processes
could contribute to producing the ions of the inner source: (a)
Evaporation from interstellar grains when they come near to the Sun;
(b) evaporation from interplanetary grains, i.e. debris from asteroids
and comets; (c) dust released from small, undetected comets; (d)
multiply-charged solar wind ions absorbed by grains and desorbed again
to form singly charged ions. Since the strengths of the partial
sources (b), (c) and (d) ought to decrease with increasing latitude,
it is likely that source (a), evaporation from interstellar grains,
contribute most of the ions at high latitude. Thus we may expect that
a more complete study of the latitudinal dependency of the ${\rm
C}^+$, ${\rm N}^+$, ${\rm O}^+$ fluxes and velocity distributions by
Ulysses will provide an unambiguous identification of the ion
population that is produced by evaporation from interstellar grains in
the inner heliosphere. This would provide a way to estimate the
content of semi-volatile components in these grains.

\section{Interaction of Interstellar Dust Grains with the
Heliosphere }\label{exclusion}
In order to compare the gas-to-dust mass ratio values, $R_{\rm g/d}$,
as determined inside the Solar System (\S \ref{totaldens}) against the values found for the LISM (\S \ref{licgas2dust}), 
we must first understand the transversal of these grains across the
heliopause region, and after that through the solar wind.
Previous treatments of this topic include, e.g.,
Levy and Jokipii (1976), Wallis (1987)  and Kimura and Mann (1998).
Comparisons between interstellar dust grains observed within the heliosphere 
with the incident interstellar population provides
a unique opportunity to tie together the properties of the interface
between the solar wind dominated heliosphere and the surrounding
interstellar cloud, the cloud itself,
and the interstellar radiation field generated by stars in the LISM and beyond.  The
charging of the grains due to processes in the LIC determines how 
effectively the grains couple to the magnetic field as the field is swept into 
the heliospheric bow shock, compressed and then swept around
the nose of the heliosphere. The structure of the field in the region between 
the heliopause and
the bowshock and, in particular, the field strength, determine how
grains with a given charge-to-mass ratio will propagate. The far ultraviolet
(FUV) radiation 
field, in particular that part of the field above 8 eV, plays a critical role in
determining the equilibrium charge of the grains (as detailed below). 

Grains which succeed in crossing the heliopause encounter the
solar wind, where they are subject to Lorentz-force interactions as
well as gravity and radiation pressure.
Thus, the {\it in situ} observed dust grain size distribution will depend
on the initial grain size distribution in the LIC, 
on the dynamics of grains injected into the heliosphere
bow shock and heliosheath, and on grain charging rates inside
and outside of the heliosphere. 
The interstellar magnetic field, FUV radiation field intensity, and 
solar cycle dependent solar wind properties all contribute to the ability of
interstellar dust grains to penetrate the heliosphere,
so that comparisons between {\it in situ} data and predicted incident
grain mass distributions should provide important constraints on all properties. 

\subsection{Heliosphere Model }\label{mhdmodel}
Dust grain traversal of the heliopause region must be determined
using a heliosphere model, which includes both neutral ISM and an
interstellar magnetic field.
The self-consistent inclusion of neutral interstellar hydrogen into models
describing the interaction of the solar wind with the LIC changes the
structure of the heliosphere when compared to heliosphere models based on purely gas dynamical or MHD
assumptions. Specifically, the neutral hydrogen mediated heliosphere is typically
of much smaller extent than a non-mediated heliosphere
because neutral gas is more compressible than plasma.
This can be seen when the gas-kinetic models of Baranov et al. (1998)
are compared with gas-dynamic models.
To date, several models
have been advanced which describe the two-dimensional (2D, Baranov and Malama 1993,
Pauls et al. 1995,  Zank et al.\ 1996) and three-dimensional (3D) Pauls and Zank
(1997a) structure of the heliosphere while retaining the dynamical influence of
the interstellar neutral H component. 
Pauls and Zank (1997b) have developed a ``convected field''
model based on their 3D hydrodynamic models. Linde et al. have 
developed an MHD gas-dynamic model (\cite{linde}).
We use a barely subsonic multi-fluid model for the heliosphere-ISM iteration.
We use the Pauls and Zank (1997a,b) and Zank et al.\ {1996}
models of the global heliosphere
to make (i) a simple estimate of the grain sizes that will be excluded from
the heliosphere, and (ii) use the observed lower limit on {\it in situ}
grain size (\S\ref{massdis}) to
estimate bounds for the LISM magnetic field strength. By balancing the length
scale of the heliospheric boundary layer, in this case the separation distance
between the bow shock and heliopause, with the Larmor radius of the dust
grains, we can determine which dust grain sizes are likely to be swept around
the heliosphere by the diverted LISM plasma component flow. Particles with gyroradii in excess of
the boundary layer thickness will enter the heliosphere relatively easily owing
to their weaker coupling to the interstellar plasma on these length scales. 
This comparison yields an estimate of the sizes of the excluded grains.
It will be seen that these results do not change significantly when we use
an MHD model which incorporates a weak interstellar magnetic field.

By assuming a LISM proton temperature $T$ = 7500 K, a (supersonic) plasma inflow
velocity of 26 km s$^{-1}$, and proton number density of 0.1 cm$^{-3}$, together with
a neutral H interstellar density of 0.14 cm$^{-3}$, the 3D simulation of 
Pauls and Zank (1997a) yields a two-shock heliosphere, i.e., one in which a
bow shock decelerates the impinging LISM flow and a solar wind termination
shock, both of which are separated by the heliopause. While the temperature
used in this model deviates from the best value (Table \ref{licprop}),
these results are insensitive to that difference.  The distance to the
termination shock is $\sim 95$ AU at the nose and the distance to the bow shock
is $\sim 235$ AU and the separation distance between the bow shock and
termination shock is $\sim 140$ au. 
These figures can vary a little depending on
parameters, and a typical separation length between the bow shock and the
heliopause for the various two-dimensional (2D) simulations is $\sim 150$ au. 
The inclusion of an interstellar and
interplanetary magnetic field (Washimi and Tanaka 1996, Pauls and
Zank 1997b, \cite{linde}) leads to a substantial increase in the strength of the magnetic
field from the LISM to the heliopause. The weak perpendicular bow shock
(compression ratio $r \sim 2$) approximately doubles the downstream magnetic
field strength. In the heliosphere model used here, the LISM magnetic field increases 
nearly tenfold in an almost linear fashion to peak at the
heliopause where it meets the interplanetary field. The interplanetary field is
compressed at the termination shock, and also increases rapidly between the
termination shock and the heliopause.
For comparison, an MHD heliosphere model which directly
incorporates an interstellar magnetic field and
allows for field compressibility between the bow shock and heliopause 
gives an overall increase of the interstellar magnetic field
between the heliopause and bow shock of about five for the
assumed low density interstellar field (1.5 $\mu$G) (Linde et al. 1997).
In this case, the relative distances of the termination shock,
heliopause, and bow shock for the model adopted here are
120 au, 200 au, and 300 au.
For a weakly magnetized ISM, the undisturbed interstellar field
jumps weakly by about a factor of 2 at the bow shock, and rises
to about 7.5 $\mu$G at the heliopause.

\subsection{Grain Charge and Gyroradius}\label{graincharge}
The dust charge may be expressed as 
\begin{eqnarray}
Q = e Z_{\rm gr} = 4\pi \varepsilon_0 a_{\rm gr} U_{\rm eq}\label{eq-charge},
\end{eqnarray}
where $U_{\rm eq}$ denotes 
the equilibrium grain surface potential, $a_{\rm gr}$ the radius 
of the (assumed) spherical grain, and
the permittivity $\varepsilon_0=8.859\cdot 10^{-14}\ {\rm C}\ {\rm V}^{-1}\ {\rm cm}^{-1}$.
Thus the Larmor radius $l_{\rm gyr}$ of the
grain depends on the grain radius quadratically,
\begin{equation}
l_{\rm gyr} = \frac{m_{\rm gr}~ v_{\rm d}}{Q ~ B} 
 =  \frac{\rho_{\rm gr} a^2_{\rm gr} v_{\rm gr} }
{3 \varepsilon_0 U_{\rm eq} B }\label{eq-gyro}.
\end{equation}
In (\ref{eq-gyro}), $B$ is the magnetic field, $\rho_{\rm gr}$ denotes the density of
the dust grain and $v_{\rm d}$ is the grain velocity perpendicular 
to $B$. 
Guided by the results of detailed numerical calculations of the
heliospheric boundary layer, we can express the variation of $B$ 
between the bow shock and heliospause (to leading order 
along the nose) as, $B = Cx + r B_{\rm ISM}$, where 
$0 \leq x \leq 150$ AU, $B_{\rm ISM}$ is the LISM magnetic field strength, 
$C \equiv (10 - r) B_{\rm ISM} / 150$ is a constant, and $r$ is the 
compression ratio at the bow shock. This expression contains the result
that $B$ reaches $\sim 10 B_{\rm ISM}$ at the heliopause, but 
MHD calculations which include field compressibility indicate maximum
compression factors of $\sim$5 (Linde et al. 1997).  
A typical value for $r$, given our assumptions is $r \approx 2$.
The grain density, $\rho_{\rm gr}$, that we assume is $2.26$ g cm$^{-3}$, appropriate for graphite grains. The speed of the Sun relative to the 
surrounding interstellar cloud, and thus the inflow speed of 
dust impinging on the heliosphere, is determined from 
observations to be $v_{\rm gr} \simeq 26$\,km\,s$^{-1}$, and we assume
$v_{\rm d}$=$v_{\rm gr}$ since the interstellar magnetic field direction is approximately
perpendicular to the heliocentric flow vector.
(The deflection of interstellar dust grains by magnetic
fields in the heliosheath regions and heliosphere is
discussed in detail by Landgraf, 1998.)
We will return to a discussion of grain charge and gyroradius in Section \ref{dynsep}
where the dynamic separation of interstellar dust and gas in the LISM
is discussed.

The primary processes responsible for charging grains in the warm
($T\sim 10^4\,$K) ISM are electrons and protons hitting and sticking 
to grains and photoelectric charging by photons from the FUV 
background. Ejection of secondary electrons due to electron and 
proton impacts are a minor effect at the temperature of the LISM,
increasing the equilibrium potential (making it less negative) by 
$\sim$5\% in the absence of photoelectric charging. To determine
$U_{\rm eq}$ we need to solve the equation for equal positive
and negative grain charging rates 
as required by grain charge equilibrium, and relate those rates to
the FUV flux and the grain potential (Draine \& Salpeter 1979). We may 
write this equation as:
\begin{equation}
g(\phi) = \sqrt{\frac{m_e}{m_p}}
\frac{s_{\rm p}^{\rm eff}}{s_{\rm e}^{\rm eff}} g(-\phi) +
\left(\frac{2 \pi m_e}{k T}\right)^{1/2}\frac{1}{n_e}J_{\rm FUV} \label{chg_rate}
\end{equation}
where
\begin{equation}
g(x) \equiv \cases{e^x, &if $x < 0$\cr (1 + x),&if $x \ge 0,$\cr}
\end{equation}
and $\phi$ is the grain potential parameter, $\phi \equiv \frac{e U}{k T}$;
$s_{\rm e}^{\rm eff}$ ($s_{\rm p}^{\rm eff}$) is the effective 
sticking coefficient for electrons (protons) including the reduction
(increase) due to secondary electron emission; $n_e$ is the electron
number density in the LISM and $J_{\rm FUV}$ is the
photoelectron current per unit grain area which depends on both the
grain size and the FUV background flux averaged over the photoelectric 
yield. Ignoring the weak dependence of
$s_{\rm e}^{\rm eff}$ and $s_{\rm p}^{\rm eff}$ on grain size, all of
the dependence on the background FUV flux, the electron density and the 
grain size can be collected in the parameter $\zeta$ (McKee et al.\ 1987):
\begin{equation}
\zeta \equiv \frac{G_0}{n_e}\left(\frac{a_{\rm gr}}{0.01\mu m + a_{\rm gr}}
\right)
\end{equation}
where $G_0$ is the yield-averaged FUV background normalized to Draine's
(1978) standard background. (As we discuss below, we now have reason to
believe that $G_0$ is less than one). 
Thus by specifying $\zeta$ we
can solve equation (\ref{chg_rate}) numerically to determine $\phi$ and
thus the charge on the grains, $Z_{\rm gr}$.

\begin{figure}[ht]
%\figurenum{gyroradii}
\epsscale{0.8}
\plotone{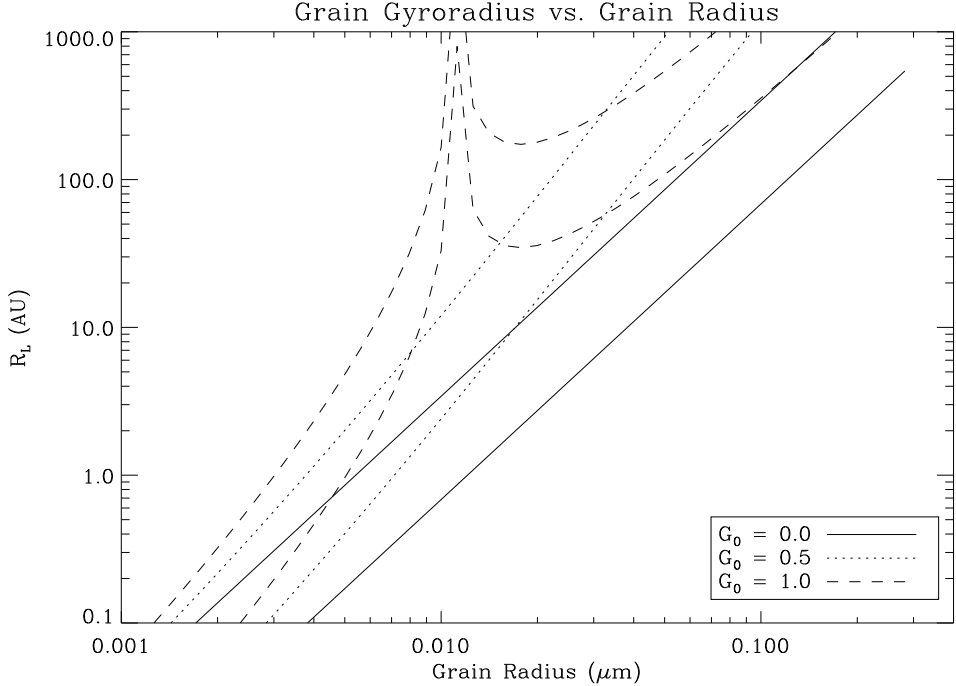}
\caption{\it Gyroradius vs.\ grain size for interstellar dust flowing
into the heliosphere for an assumed ISM field of 1.5\,$\mu$G and
for different strengths of the FUV background parameter, $G_0$. The
upper curve for a given $G_0$ corresponds to the gyroradius at the
bowshock and the lower curve is the gyroradius at the
heliopause.\label{gyroradii}}
\end{figure}

We take the electron density in the Local Cloud to be 0.1\,cm$^{-3}$
(\S\ref{LIC_model} and Table \ref{licprop}). With this choice, and a particular choice for $G_0$,
we can then determine $U_{\rm eq}$. Expression (\ref{eq-gyro}) is then a 
relation between $l_{\rm gyr}$ and $B_{\rm ISM}$. In Figure \ref{gyroradii} we
adopt a canonical LISM value for $B_{\rm ISM} = 1.5\,\mu$G (Table \ref{licprop}), 
and plot the dust gyroradius as a function of 
$a_{\rm gr}$ across the heliospheric boundary layer. For a given value
of $G_0$ (as noted in the figure), the range in gyroradius from 
bowshock (upper curve) to heliopause (lower curve) is shown in the figure.
The increase in Larmor radius with increasing FUV flux is due to the
decrease in the absolute value of the potential.  As the FUV flux
increases, the photoelectric current increases and the potential, which
in most cases is negative, 
becomes more positive. 

%\placefigure{gyroradii}

The sharp upturn of the Larmor radius at $\sim 0.011\,\mu$m for the 
$G_0 = 1$ case is due to the crossing from negative charge for smaller 
grains to positive charge for larger grains.  Note that further
increases in the FUV flux ($G_0 > 1$) would continue to pull up $l_{\rm gyr}$
for small grains which were still negatively charged, but would pull
down $l_{\rm gyr}$ for the large grains as their positive charge would
increase.  Formally, for large enough FUV flux 
($G_0/n_e > \zeta_{\rm crit}$) there is a grain size at which grains 
are neutral. Though it is unlikely that any significant number of 
inflowing grains will in reality be completely neutral, these results 
do point up the intriguing possibility that for a small range in grain 
size, the Larmor radii will be quite large. Thus it is possible that
a particular size 
range of interstellar grains could be significantly overabundant in 
the heliosphere. A detection of such a signature in the data would
give us direct evidence on the charging of the dust, would tightly
constrain the FUV flux in the Local Cloud and would allow much more
reliable inferences for the large end of the dust size distribution.

\subsection{Exclusion of Small Particles at the Helio\-pau\-se}\label{hpfiltration}
Figure \ref{gyroradii} clearly demonstrates that grains smaller than
$\sim 0.01\,\mu$m will be excluded from the heliosphere under most
conditions. It is also evident that gyroradii of interstellar grains are
strongly dependent on the strength of the FUV background. Taking $G_0 = 0.5$ as a reasonable value, we find that for grains of radius 
$a_{\rm gr} = 0.04\,\mu$m, the gyroradius is reduced from $\sim 500$ AU at
the bow shock to $\sim 100$ AU at the heliopause. Thus, even for a relatively
weak LISM magnetic field of 1.5\,$\mu$G, dust grains whose radius is less 
than $\sim 0.05\,\mu$m are likely to be excluded from the Solar System since 
they are sufficiently strongly coupled to the LISM plasma that they 
will tend to be diverted around the heliosphere. 
This exclusion limit is comparable to the lower limit of the mass sensitivity of the
Ulysses and Galileo detectors (\S\ref{massdis}), and indicates that 
increasing the sensitivity of dust detectors on interstellar
spacecraft will yield valuable science. 

\begin{figure}[ht]
%\figurenum{BISM}
\plotone{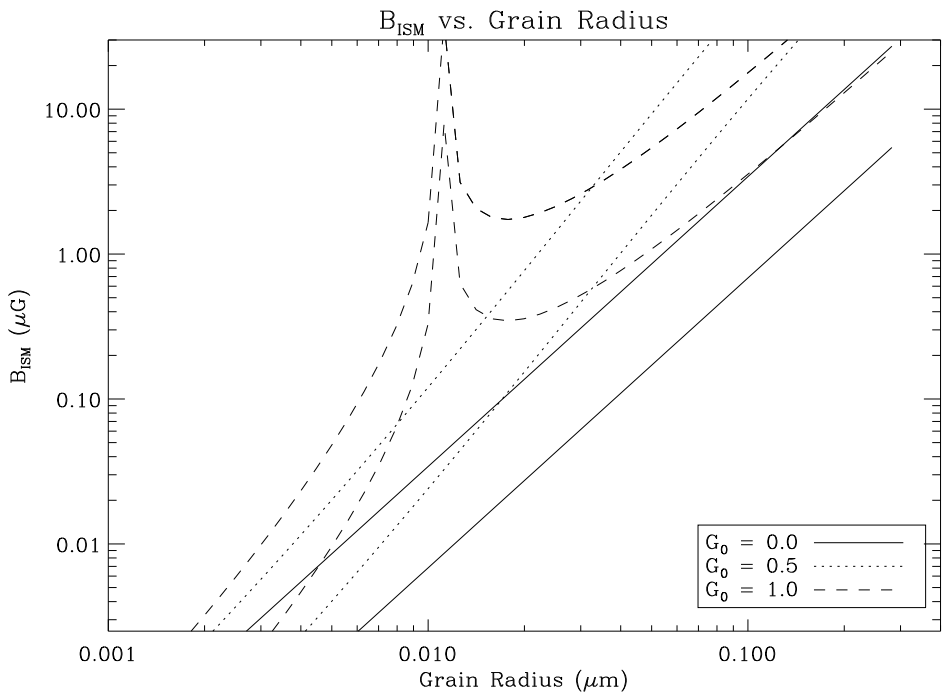}
\caption{\it Interstellar magnetic field vs.\ grain size for
an assumed Larmor radius of 150\, AU and
for different strengths of the FUV background parameter, $G_0$.
We assume that grains with gyroradii less than 150\, au, the approximate
distance between the bowshock and the heliopause, will be excluded from
the heliosphere. The upper curve for a given $G_0$ corresponds to the
field if the grains are excluded at the bowshock and the lower curve
is for exclusion at the heliopause.\label{BISM}}
\end{figure}

If one now turns expression (\ref{eq-gyro}) around by balancing $l_{\rm gyro}$ against
the boundary layer scale length, one determines a relation between the
interstellar magnetic field $B_{\rm ISM}$ and the grain radius, from which 
one can infer limits on the strength of the LISM magnetic field by 
observing which dust grains are excluded from the heliosphere. By 
adopting a separation length between the bow shock and heliopause of 
150\, AU (\S\ref{mhdmodel}), we obtain Figure \ref{BISM} in which $B_{\rm ISM}$ is
plotted as a function of $a_{\rm gr}$. The range of $B_{\rm ISM}$ 
values, which is a function of the variation of the magnetic
field strength between the bowshock and heliosphere, is 
constrained by the pairs of lines for each value of $G_0$ as before.
If dust grains of radius $0.04\,\mu$m are to be excluded from the 
heliosphere, then (again assuming $G_0 = 0.5$) the strength of the 
LISM magnetic field has to lie in the range 1--5\,$\mu$G. The upper 
bound is in fact remarkably close to the estimated upper
limit obtained by Gloeckler et al.\ (1997), from comparing pick-up ion data
with heliosheath models.
Arguments such as these can be used to derive the interstellar
magnetic field from {\it in situ} detections of low mass 
interstellar grains in the heliosphere regions with future interstellar
spacecraft.

%\placefigure{BISM}

\subsection{Heliosphere--ISM Transition Region and Dust Processing Timescales 
in LISM }\label{transition}
The destructive processes that could operate on interstellar dust 
during its transport through the heliosheath regions and into the 
Solar System are thermal and inertial 
sputtering, and the grain-grain collisional processes of vaporization 
and shattering, similar to destruction processes in the LISM.
these processes are insignificant at the low LIC densities, however.
The threshold temperatures for the thermal sputtering of all likely grain 
materials are in excess of 100,000 K (Tielens et al. 1994), and thus 
will be unimportant at LISM temperatures of $\sim$7000 K. 
Likewise, in the transition through the relatively weak bowshock and through 
the heliopause the temperatures (T$<$30,000 K typically) are below the thermal sputtering threshold, 
and this process can therefore be ignored. 

The threshold velocity for inertial sputtering, the process of grain erosion in gas-grain collisions, 
is of order 30 km s$^{-1}$. Thus, if the inflowing interstellar gas is 
rapidly slowed at the bowshock and the heliopause, and the larger momentum 
of the grains causes them to drift through this almost stationary gas, then 
inertial sputtering will be unimportant for the inflow velocity of 26 km s$^{-1}$. 
However, in these regions the dust will be betatron accelerated with respect 
to the gas, but the velocities (for LISM magnetic fields of 
order 5 $\mu$G) will not result in significant inertial sputtering for  
the distances, densities and velocities appropriate to the transition from 
the LISM to the Solar System. For instance, in the more extreme environment 
of a 50 km s$^{-1}$ interstellar shock wave in the warm ISM, with a density the same as that 
in the LISM, inertial sputtering results in the loss of only 0.1\% of the 
grain mass, independent of grain  radius (Jones et al. 1996).
Again using the 50 km s$^{-1}$ shock data (Jones et al. 1996) as an
illustration it can also be shown that the effects of grain shattering
in grain-grain collisions will be minimal in the bowshock and
heliopause regions. Thus, other than filtering out the smaller grains,
it is unlikely that there is significant physical processing of the
dust as it traverses the bowshock and heliopause regions.

\subsection{Dynamics of Interstellar Dust Particles inside the
Heliosphere}\label{hsfiltration}
After the grains have penetrated the heliopause region, they enter the
heliosphere in which they experience gravity, radiation pressure and
the Lorentz-force caused by the solar wind magnetic field and their
equilibrium charge in the heliosphere (surface-potential $U_{\rm eq}$=+5 V)
(\cite{morfill85}). Due to the low density, the drag force by solar wind
ions is negligible (\cite{gustafson96}). The relative strength of these
forces depends in first approximation on the grain size, which
determines the radiation pressure efficiency $\beta$
(\cite{gustafson96}) and the charge-to-mass ratio $Q/m_{\rm gr}$ (\cite{gruen96a}).

For large grains ($a_{\rm gr} \geq 1\ {\rm \mu m}$), the Lorentz-force can be
neglected. They are on hyperbolic Kepler-orbits. Gravitational
focusing enhances their spatial density downstream of the Sun, whereas
the density upstream of the Sun is very close to the value in the LIC.
For grains with sizes of $\approx0.4\ {\rm \mu m}$ radiation pressure
dominates gravity and the grains get repelled by the Sun. Even smaller
grains with $a_{\rm gr}\approx 0.1\ {\rm \mu m}$ are dominated by the
Lorentz-force (\cite{levy76}), which is in good approximation given by
$\vec{F}_{\rm L} = Qv_{\rm sw}B_{\phi}\vec{e}_\theta$, where $Q$ is
the grain charge (Eq. \ref{eq-charge}), $v_{\rm sw}$ is the velocity of the solar wind
particles, $B_{\phi}$ is the azimuthal component of the solar wind
magnetic field and $\vec{e}_\theta$ is the normalized electric vector in polar
direction. The large scale interplanetary magnetic field changes its
polarity with the 22-year solar cycle. During one half of the solar
cycle the Lorentz-force points towards the ecliptic north in the
northern hemisphere and to the south in the southern hemisphere. Thus,
the grains are defocused from the plane of the solar equator. In the
second half of the solar cycle, the field-polarity is reversed and
thus the particles in the northern hemisphere are deflected to the
south and particles on the southern hemisphere are deflected to the
north.

The solar cycle that started after the solar maximum around $1991$ is
defocusing until the next solar maximum in $2002$. In both phases of
the solar cycle very small grains experience a net deflecting
Lorentz-force, because the magnetic field is at rest with respect to
the frame of the solar wind.  The size of the smallest grain which can
penetrate to a given radial distance from the Sun depends on the solar
cycle phase.  The size-dependent filtering of interstellar grains within the heliosphere, which
governs the size distribution of interstellar grains inside the
heliosphere, depends on the total strength of the solar wind magnetic
field and its perturbations at low latitudes and the equilibrium
charge induced on the grains. In first approximation the filtration is
independent of the solar wind velocity because the strength of the
azimuthal component of the magnetic field decreases when the wind
velocity increases leading to a constant Lorentz-force
(\cite{gustafson79}).  The effect of filtration on the {\it in situ}
mass distribution above the detection mass threshold
(Figs. \ref{massdis_fig} (a), (b) and Fig. \ref{densdis}) is that
submicron-sized grains are depleted with respect to bigger grains,
explaining the relative absence of 10$^{-15}$ to 10$^{-13}$ g grains
compared to the MRN distribution.

In an independent investigation of this problem, Kimura and Mann (1998) find
similar results.  They find that the influence of Lorentz forces on grains
with masses less than 10$^{-16}$ g (corresponding to $a_{\rm gr}$$\sim$0.02
$\mu$m for density 2.5 g cm$^{-3}$) cause significant deflections of the
grain propagation directions from the original flow.

\section{Dust Grains in Local ISM}\label{LICdust}
The diffuse ISM is an inhomogeneous environment (see, e.g.,
the review \cite{dorschhen}, or the array of diffuse clouds identified towards
23 Ori, Welty et al. 1999).  In this section we tacitly assume
a classical view of dust in the interstellar cloud surrounding the Solar System
- i.e. that the dust grains of all sizes are formed by condensation from
interstellar gas of cosmic composition.

With this assumption we subsequently use observations of interstellar absorption lines in 
the spectra of nearby stars
to determine the gas-to-dust mass ratio, $R_{\rm g/d}$, for the cloud
feeding interstellar dust grains into the Solar System.
This derivation of $R_{\rm g/d}$ requires two assumptions: first - that a
nominal reference abundance describes abundances in the ISM
and also that the total mass of the atoms ``missing'' from the gas phase
is equal to the total mass of the interstellar dust grains.
This standard method of determining the gas-to-dust mass ratio 
(e. g. \cite{ss}; \cite{meyerAIP}) stands in contrast
to values determined by modeling reddening and extinction curves,
and scattered light (e.g., \cite{spitzer78,whittet}).
For example, the widely-adopted Draine and Lee (1984) dust model
which models extinction curves requires 300 ppm of carbon in 
graphite grains, and approximately solar
abundances of Si, Mg and Fe. This results in a gas to dust ratio of
$\sim 110$ with respect to H  (or $\sim 160$ with respect to H+He).
Large grains are not coupled to the gas (\S\ref{dynsep}), and both very small
and very large presolar grains from supernova
and/or circumstellar envelopes of active giant stars are seen in
meteorites (\S\ref{hoppe}).
If presolar type grains are mixed in with, or seed, the 
interstellar grain population, the basic assumption of this section
is invalid.  In other words, the dust mass can not be assumed
to be equal to the mass of atoms ``missing'' from the
gas phase, when the ``missing mass'' is calculated by comparing the total masses
of the observed gas and hypothetical reference abundance gas.

The only direct evidence for interstellar dust grains in the cloud
surrounding the Solar System is the optical polarization of light
from nearby stars.  These data are discussed in Appendix \ref{polarization}.
Because of the low total column densities, the polarization data do not directly constrain
grain properties (e. g. Frisch 1991), but indicate silicate grains
may be present (\S\ref{licgrainsize}).

The abundances of refractory elements in the nearby ISM are enhanced with 
respect to refractory abundances in cold interstellar gas (Frisch 1981).  
In Section \ref{lism_abs}
we show that these refractory depletions are relatively well behaved.
Comparing LIC depletions with the range of Mg and Si depletions
found by Fitzpatrick (1997), it is seen that the cloud around the solar
system is a weakly depleted cloud, as confirmed by GD.
An early model for the LIC proposed that the enhanced abundances of
refractory elements,
when compared to more distance cold clouds, are from the expansion
of a shock front associated with the Loop I supernova remnant
to the solar location (\cite{fr81,fr95}).  The destruction of LIC
dust grains by a shock is modeled in Section \ref{SF}.

\subsection{Gas-to-Dust Mass Ratio in the LISM}\label{licgas2dust}
Does the gas-to-dust ratio $R_{\rm g/d}$ found from the {\it in situ} data
agree with the value inferred from astronomical observations of the LISM?
The gas-to-dust ratio in nearby interstellar gas is determined here for
two sightlines through the LISM gas ---
LIC velocity component towards $\epsilon$ CMa and the warm
neutral cloud complex towards $\lambda$ Sco (which includes the LIC).  
These values are also compared to values for more distant diffuse
gas seen towards 23 Ori.  Total column densities towards
$\epsilon$ CMa are an order of magnitude below values towards $\lambda$
Sco (and 23 Ori), therefore in principle the LIC velocity component towards $\epsilon$ CMa
(at --17 km s$^{-1}$, heliocentric velocity)
should provide the best measure of abundances relevant to LIC dust grains.  
Together observations of gas-phase elements in these clouds
constrain the dust grain mass and composition by comparison
with a hypothetical total ``reference'' abundance of the element in the combined gas 
plus dust phases.  
We assume the difference between the mass of the gas with reference 
abundances and the mass of the observed gas is equal to the mass
of the dust grains.
This method of determining the gas-to-dust ratio in the LIC depends 
on the underlying assumption that both the reference abundance and gas-dust 
mixing are uniform in the ISM.

The gas-to-dust mass ratio is determined here using the relation:
\begin{equation}
R_{\rm g/d} = \frac{{\cal M}_{\rm G }}{{\cal M}_{\rm R'} - {\cal M}_{\rm G'}}~=~
\frac{\sum_{i,gas}   n_{\rm g}({\rm X_{\rm i}})  A_{\rm i} }{ \sum_{i,dust}   n_{\rm d}({\rm X_{\rm i}})  A_{\rm i}} ~=~
\frac{\sum_{i,gas}   {\rm PPM}_{\rm g}(X_{\rm i})  A_{\rm i} }{ \Sigma_{i,dust}   {\rm PPM}_{\rm d}(X_{\rm i})  A_{\rm i}}.\label{g2d}
\end{equation}

The quantity ${\cal M}_{\rm G}$ is
the total mass of all elements in the gas phase including He, and
${\cal M}_{\rm G'}$ is the total mass in the gas phase of the elements excluding H and He.
Note that He contributes $\sim$29 \% of ${\cal M}_{\rm G}$.
The quantity ${\cal M}_{\rm R'}$ is the total
mass of all elements heavier than He in the hypothetical reference abundance
gas.  
If all elements heavier than He were condensed onto the dust grains, the 
total mass of the dust grains would be
${\cal M}_{\rm R'}$ (yielding gas-to-dust mass ratios of
64 and 101, respectively, for the solar and B-star abundances given
in Table \ref{tab-ppm}).  Since
H and He are not expected to be incorporated into the grains, the
denominator is summed over elements heavier than He, while the numerator
summation is over all elements (although heavier elements contribute relatively little mass to the gas).  Here,
only those elements are considered which have solar 
PPM abundances greater
than 18, namely C, O, N, Si, Mg, Fe, and S.
The quantity $ n_{\rm g}({\rm X_{\rm i}}$) is the space density in the gas phase of element X$_{\rm i}$,
$n_{\rm d}({\rm X}_{\rm i})$ is the space density in the dust phase of element X$_{\rm i}$, and $A_{\rm i}$ is the atomic mass of element i.
The quantity ${\rm PPM}({\rm X_i})$~=$\frac {N({\rm X_i})}{N({\rm H})} ~ 10^6$, with ${\rm PPM}_{\rm g}({\rm X_i})$, ${\rm PPM}_{\rm d}(X_{\rm i})$ representing the
number of atoms in ``parts per million'' in the gas and dust, respectively.  The quantity $N({\rm X_{\rm i}})$ is the
column density of element $X_{\rm i}$.  
Uncertainties in the correct reference abundances to use for interstellar material
are the major error in determining the gas-to-dust mass ratio in 
nearby ISM.  
Grain mass is dominated by a combination of the most abundant and heaviest elements.

%\placetable{tab-Rgd}
\input{table_g2d}
%\placetable{tab-ppm}
%\input{tab-ppm}

The gas-to-dust mass ratio calculated for the LISM is highly sensitive to the assumed reference abundance pattern. 
Several recent studies find that the correct reference
abundances for the ISM are $\sim$70\% of solar values
(Snow and Witt 1995, 1996; \cite{fitzFeMnCr,fitzMg,meyerC,meyerN,meyerKr}). 
The strongest observation-based argument for using B-star reference abundances for the ISM is based on observations
of interstellar Kr, which has the dominant form Kr$^{\circ}$ in the ISM (the ionization
potential is 13.99 eV), and which as a noble element should not be
depleted onto dust grains (Cardelli and Meyer 1997).  Cardelli and Meyer found that the interstellar
Kr$^{\circ}$ abundance does not vary from sightline to sightline, and that
it is independent of the fraction of hydrogen present in molecular form; they derived an interstellar Kr abundance in the solar vicinity of $\sim$60\%
of the Solar System value.
Observations of another undepleted element, N, in the form of N$^{\circ}$, give N abundances $\sim$80\% solar.  
Uncertainties in the diffuse cloud ionization of N (with an ionization potential slightly
above that of H), and
in underlying oscillator strengths, however, render this result inconclusive (\cite{meyerN}). 
Observations of O$^{\circ}$ and C$^{+}$, where both elements are grain constituents, suggest reference abundances $\sim$67\% 
and  $\sim$60\% solar, respectively in interstellar gas (\cite{meyerO},
\cite{meyerC}; \cite{snowwitt}).   The gas-to-dust ratios in the LIC
will be calculated for both assumed B-star and solar reference abundances.

The calculation of elemental abundances requires knowledge of H$^{\rm o}$ column densities, which for the case of the LIC component towards
$\epsilon$ CMa is somewhat uncertain because of strong stellar L$\alpha$
absorption lines.  Despite these uncertainties, we will show in the
following paragraphs that the high values for $R_{\rm g/d}$ found for
the LIC component towards $\epsilon$ CMa are typical of warm
neutral clouds, suggesting that the H$^{\rm o}$ column density uncertainties
do not yield misleading conclusions.
Three independent types of arguments indicate that
the H$^\circ$ column density of the LIC component towards $\epsilon$ CMa
has a value in the range log $N$(H$^\circ$)=17.3-17.5 cm$^{-2}$.
Trace element measurements, including data on undepleted S,
yield values log $N$(H$^\circ$)=17.30 cm$^{-2}$ if solar reference abundances
are assumed, and log $N$(H$^\circ$)=17.51 cm$^{-2}$ if B-star reference
abundances are assumed (\cite{dupin98,gry95,gry96,gry98}, collectively referred to as 
GD; Table \ref{tab-Rgd}).  These low column densities are reinforced by
observations of low column densities towards Sirius
($\alpha$ CMa), which is within
$\sim$5$^{\rm o}$ of $\epsilon$ CMa but only 2.7 pc from the Sun.
The Sirius data give log $N({\rm H^o})$=17.23$^{+0.17}_{-0.28}$ cm$^{-2}$
for the LIC component (\cite{bertin95}).  The third support for low
LIC column densities towards $\epsilon$ CMa are models of the $\epsilon$ CMa
extreme ultraviolet spectrum, corrected for transmission through an interstellar
cloud medium with solar abundances.
These models require log $N$(H$^\circ$)$<$17.70 cm$^{-2}$ for the total H$^\circ$ column density (which is greater than the LIC cloud column density,
\cite{aufd}).

Values for $R_{\rm g/d}$ for the LIC component in the star $\epsilon$ CMa
and the LISM gas in $\lambda$ Sco are given in Table \ref{tab-Rgd}.
These values are based on equation \ref{g2d}, and the interstellar
absorption line data presented
in Appendix \ref{LICcolumns} (which are based on data from
\cite{dupin98,gry95,gry96,gry98,york83}).
For $\epsilon$ CMa, R$_{\rm g/d}$=427$^{+72}_{-207}$ and 551$\pm^{+61}_{-251}$
for assumed solar and B-star abundances, respectively.  
For $\lambda$ Sco,
R$_{\rm g/d}$=137$^{+16}_{-40}$ and $406^{+58}_{-243}$
for assumed solar and B-star abundances, respectively.  
For comparison, for the diffuse interstellar ``WLV" component
found by Welty et al. (1999) towards 23 Ori,
R$_{\rm g/d}$=127$^{+13}_{-12}$ and $399^{+126}_{-120}$
for assumed solar and B-star abundances, respectively.
The 23 Ori WLV component has $\sim$0.3 dex more H$^\circ$
than $\lambda$ Sco, and shows similar values for R$_{\rm g/d}$.

The uncertainties quoted here are based on
uncertainties in the input gas column densities.
For the $\epsilon$ CMa data, the uncertainties represent the widest range 
of profile-fits yielding acceptable fits for the range of absorption lines 
under consideration (C. Gry, private communication).
Although it is hard to translate the quality of results produced by
profile fitting techniques (which require judgement calls) into
root-mean-square deviations, the
quoted accuracies should be good to at least the 2$\sigma$ level,
and possibly better.  For the purposes of comparing the gas-to-dust
mass ratios derived from astronomical data versus {\it in situ} data,
we conclude that the resulting differences are real.  This conclusions is
also strongly supported by the high $R_{\rm g/d}$ values found
for the other diffuse cloud sightlines considered here.
Note that small uncertainties on input gas-phase column densities
translate into large fractional uncertainties in the dust PPM values 
for elements that are present primarily in the gas-phase 
(see Appendix \ref{LICcolumns}).

From the high values of R$_{\rm g/d}$ found in the preceding paragraph,
for three separate diffuse clouds,
when B-star abundances apply, we conclude that in this case
there is a clear disagreement with the
{\it in situ} value of 94.
Clearly for B-star reference abundances, 
there is not enough mass ``missing'' from the gas phase
to form the dust grain population observed by the spacecraft.
Assuming, instead, 
solar reference abundances does not bring the gas-to-dust
mass ratio for the LIC component towards $\epsilon$ CMa into agreement with
the {\it in situ} data, but does bring R$_{\rm g/d}$ in 
the higher column density material towards $\lambda$ Sco and 23 Ori into agreement.
Note that typical upper limits based on reddening curves in high column density sightlines
give R$_{\rm g/d}$$<$170 (where the upper limit applies because
large dust grains are not included in the calculation, \cite{spitzer78}).

The gas-to-dust ratio R$_{\rm g/d}$ calculated from the missing gas mass 
is a fundamental tool that can be used to probe the properties of dust grains in the diffuse ISM.  
In the case of the LIC, the larger values of R$_{\rm g/d}$ appear to be
indicative of grain destruction by interstellar shock fronts, as is
explored in Section \ref{SF}.  The somewhat higher column densities for
the $\lambda$ Sco and 23 Ori sightlines indicate that a blend
of components is present, and that blend probably includes clouds with
variable grain properties (e. g. \cite{weltylead,welty96,welty23Ori}).

The $\epsilon$ CMa LIC component has
no C available for the dust, a manifestation of the
``carbon crisis'' (e. g. \cite{dwek}).  Most of the dust mass is carried by
refractories, with 45-50\% of the dust mass carried
by Fe, and 65-70\% carried by Mg and Fe combined.
The lack of carbon available for the interstellar dust grains, 
and the presence of SiC presolar grains
in meteorites (\S\ref{presolar}), may indicate an
additional carbon-rich  grain source, beyond
gas condensation from circumstellar dust grains.

\subsection{Models of LISM Dust}\label{licgrainsize}
In this section, we adopt a simple core-mantle grain model to explain
the observed depletions.  Following the approach given in Savage and
Sembach (1996, SS), we use halo cloud sightline abundances to define
the grain core, and assume that material in the grain beyond that
required for the core composition must be contained in the grain
mantle.  Savage and Sembach use observations of halo stars to model
the cores of interstellar dust grains.  Subtraction of the core
composition from the total dust composition given in Appendix
\ref{LISM_abundances} (Table \ref{tab-ppm}) then yields the mantle
composition. The LIC is among the least depleted sightlines and there
is typically little mantle material on the grain cores.

For the grain core and solar abundances (based on SS halo
star results), Mg : Si : Fe (and the ratio
(Mg+Fe)/Si) are 1.1 : 0.6 : 1.0 (3.3).  For B-star abundances, these
ratios are 0.9 : 0.2 : 1.0 (8.6).  For the little grain mantle
material that remains on the dust in the $\epsilon$~CMa LIC component
the solar reference abundances give Mg : Si : Fe are 0.5 : 0 : 1.0.
Thus, it appears that the Si is to be found predominantly in the grain cores for the LIC.
These ratios do not change significantly for B-star abundances. In 
contrast, the grain mantles in the cold cloud complex towards
$\zeta$ Oph yield Mg : Si : Fe ((Mg+Fe)/Si) values of 1.4 : 2.6 : 1.0
(0.9) for abundances.  For B-star abundances, these mantle ratios are
1.6 : 2.6 : 1.0 (1.0).  
These comparisons indicate that independent of the assumed reference abundance
the LIC grains have been stripped of the silicate rich mantles 
found in cold cloud grains.  Grain compositions are summarized in Table \ref{tab-depl}, using data from Appendix \ref{ap_depl}.

%\placetable{tab-depl}
\input{tabdepl}

In the cold cloud the mantle composition is
enriched in Si (and Mg) when compared to the grain cores.  However,
grain mantles in the warm LIC cloud contain no Si and have very low
mass in comparison to the cold cloud mantles, i.e., most of the
original grain mantle material has been eroded into the gas phase.
The enhanced abundances of refractories in the gas phase of the LISM
(\cite{fr81,valCa}) imply grain destruction in shock fronts, and the
least refractory grain components in the mantle are eroded first.  The
material towards $\lambda$ Sco represents a blend of components, and
an intermediate case.  Assuming solar reference abundances, Fe
constitutes 46\%, and 17\% of the grain mass for the LIC cloud towards
$\epsilon$ CMa and the LISM towards $\lambda$ Sco, respectively.  This
indicates that there is relatively more silicate dust towards
$\lambda$~Sco than $\epsilon$~CMa, indicating less total grain
destruction consistent with the models below.  These data indicate
that cores are Fe rich compared to the grain mantles.

Given the limiting chemical composition of silicates; i.e., a ratio
(Mg+Fe)/Si $\sim 1$ implies a pyroxene-type silicate composition,
(Mg+Fe)/Si $\sim 2$ implies a olivine-type silicate composition,
(Mg+Fe)/Si $> 1$ implies a mixed oxide/silicate composition.  Using
these criteria this implies that metal oxides, in association with
some silicate material, are the components making up the dust grain
cores.  However, the fact that Fe and Mg are quite well correlated
(\S\ref{lism_abs}) in the LISM indicates that the grain mantles, which
have been eroded back into the gas phase, are silicates.

In order to satisfy the overall and sub-component structural and
compositional stoichiometry of the dust phase one can invoke a
`silicate' grain core and mantle structure with mineralogical
composition which, for the least refractory mantle component, is
independent of the reference abundance adopted because one measures
directly the composition from the gas phase abundances of the atoms
eroded from the grains. However, the grain core composition is
sensitive to the adopted reference abundance.
The following schematically illustrates the
likely dust phases and their approximate compositions:

Core -- mixed Olivine-type silicate and oxide phases: 
\begin{center}
{ (Mg,Fe)$_2$SiO$_4$/MgO/FeO$_x$ }
\end{center}

Mantle -- Pyroxene-type silicate phase:
\begin{center}
{ (Mg,Fe)SiO$_3$ }
\end{center}

Studies indicate that the silicon in interstellar dust seems
to be preferentially incorporated into a less refractory mantle phase,
and that the metals Mg and Fe are generally in the more resilient core
(e.g. \cite{ss,tielens98}).  The good correlation between Mg and Fe
column densities in the LISM reinforces this conclusion (next
section).  For the LIC cloud, the paucity of Si in dust mantles
indicates that the dust mantle component has mostly been eroded into
the gas. Thus, the differentiated `silicate' grain structure, which is
consistent with the observed depletions in cold clouds, will give rise to a
differential rate of return of the elements to the gas through the
effects of sputtering in supernova-generated shock waves.  Silicate
grains appear to be the main carrier of interstellar polarization, and
therefore may explain the polarization of light of nearby stars
(Appendix \ref{polarization}).  If solar reference abundances are
correct, the relatively low values for $R_{\rm g/d}$ towards $\lambda$
Sco indicate a mixture of clouds where at least one cloud contains
grains which have retained their silicate mantles, and
therefore polarize starlight.  

\subsection{Variations in LISM Abundance Patterns}\label{lism_abs}
The differences in $R_{\rm g/d}$ between LISM towards $\lambda$ Sco and
$\epsilon$ CMa and the likelihood that the LISM has been shocked (\cite{fr81},
also see \S\ref{SF}), 
suggest abundance variations may be present in the LISM gas, which is
the nearest $\sim$10$^{19}$ cm$^{-2}$ column density of gas.
In warm gas, the refractory elements Fe and Mg are sensitive
tracers of the site of grain destruction in shocks,
and therefore should be spatially correlated.
This follows from the fact that in cold clouds virtually all Fe and Mg are in grains,
so that the destruction of a few percent of the
grain mass changes gas phase abundances by large amounts.
Also, if the main contribution to observed Fe and Mg in the gas phase comes from
grain destruction, Fe and Mg should be well correlated in the LISM.  Figure
\ref{fe_mg_fig} shows this to be true, where points plotted are 
drawn from the data in Appendix \ref{ap_depl}, Table \ref{tab-femg}.
In addition, the good correlation seen between Fe$^{+}$ and Mg$^{+}$ column densities
indicates that column density uncertainties due to line saturation 
are relatively insignificant.  Previous studies of depletions give
similar results over longer sightlines (e.g., \cite{fitzMg,fitzFeMnCr}).

In section \ref{dynsep} it is shown that frictional drag couples gas and dust
poorly (scale sizes $\sim$$14$ pc for $a$=0.1 $\mu$m grains),
so that when shock fronts both transport gas and destroy grains, 
local variations in the both gas-to-dust mass ratios and refractory
abundances may result on sub-parsec scale-sizes.

%\placetable{tab-femg}
%\placefigure{fe_mg_fig}

\begin{figure}[ht]
\epsscale{0.6}
\plotone{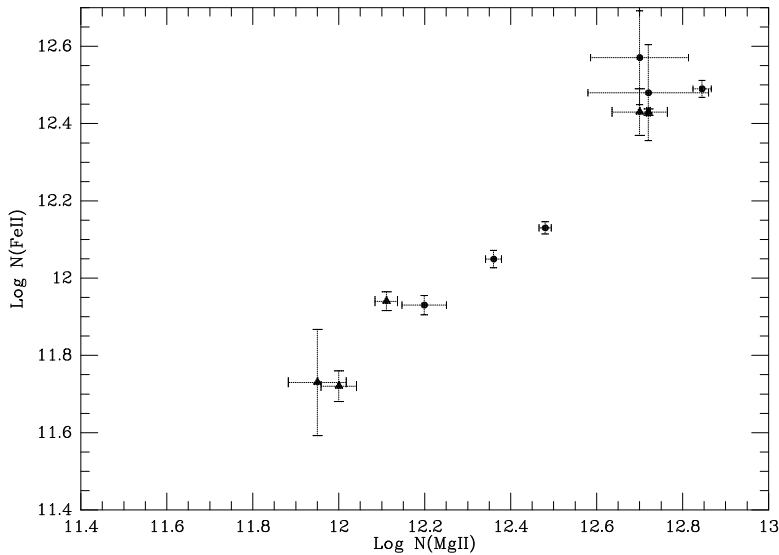}
\caption{\it Interstellar Fe$^{+}$ versus Mg$^{+}$ column densities are
plotted for nearby stars.   The data are from Table \protect\ref{tab-femg}.
Components plotted as circles or triangles indicate velocity
components that are or are not (respectively)
at the LIC cloud velocity (\S5).\label{fe_mg_fig}}
\end{figure}

In contrast, variable grain destruction in the LISM would be 
manifested by a poor correlation between Fe$^+$ (or Mg$^+$) and H$^\circ$.  
Such a poor correlation is found in Fig. \ref{fe_hI_fig}, based on 
data in Appendix \ref{ap_depl}, but an alternative explanation invoking
variable ionization would also explain this poor correlation.

%\placefigure{fe_hI_fig}
\begin{figure}[ht]
\epsscale{0.60}
\plotone{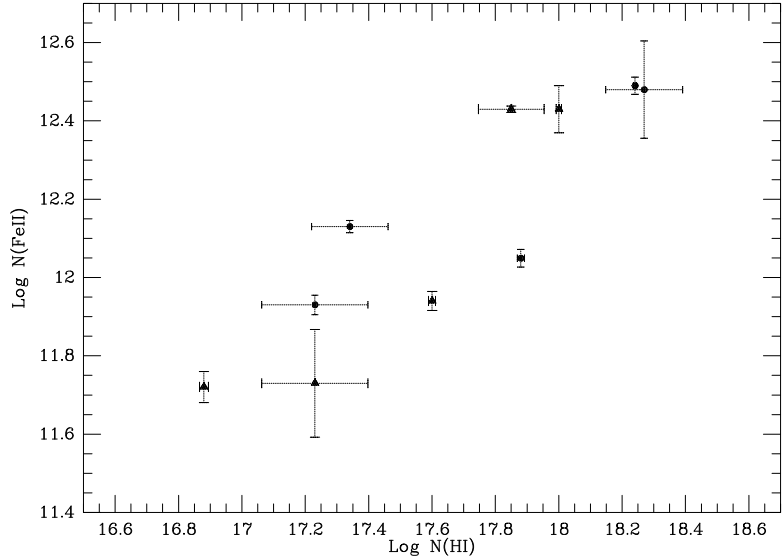}
\caption{\it Interstellar Fe$^{+}$ versus H$^{\circ}$ column densities
are plotted for nearby stars.  The data are from Table
\protect\ref{tab-femg}.  Symbols same as
Fig. \protect\ref{fe_mg_fig}.\label{fe_hI_fig}}
\end{figure}

\subsection{Shock Fronts, LIC Grain Survival and Size Distributions} \label{SF}
Simple basic models for a shocked LISM have now been presented (\cite{fy91};
\cite{lalgrz}; \cite{sonett}), but these models do not include shock
velocity-dependent grain destruction. Low velocity shocks ($\sim$20 km
s$^{-1}$ with respect to preshock gas) are used to explain velocity
gradients in nearby interstellar gas (\cite{fy91}; \cite{lalgrz}), and
a high velocity shock model has been invoked to explain enhancements
in the $^{10}$Be isotopes deposited in ice cores (V$\sim$100 km
s$^{-1}$, and a thickness of the order of 1 pc) (\cite{sonett}).

\begin{figure}
\plotone{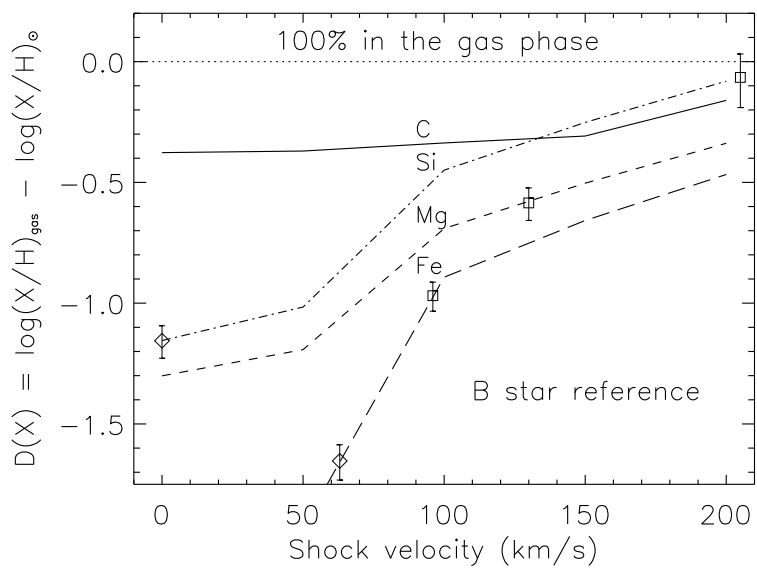}
\caption{\it Elemental depletions $\delta_{\rm H}$(X) for C (solid),
Si (dash-dotted), Mg (short-dashed) and Fe (long-dashed) as a function
of shock velocity for the B-star reference abundance core-mantle grain
model of Section 5.2.  The level of grain erosion was derived from the
results of Jones, Tielens and Hollenbach (1996). Also shown `tied' to
the depletion curves at appropriate velocities are the depletions for
the lines of sight toward $\epsilon$ CMa (squares with error bars) and
$\lambda$ Sco (diamonds with error bars), these data are taken from
Table 2.
\label{figSF}}
\end{figure}

The enhanced abundances of refractories in interstellar gas are generally
attributed to the destruction of interstellar dust grains by passing
supernova-generated shock fronts. Using the shock models and results
of Jones et al. (1994, 1996) we have calculated the depletions of the
elements C, Si, Mg and Fe for unshocked dust, and for dust subjected
to shocks of velocity 50, 100, 150 and 200 km s$^{-1}$ the results are
shown in Figure \ref{figSF}. The depletions in nearby interstellar gas can
be used to constrain the likely velocity to which nearby interstellar
material may have been shocked. Given the depletions found locally
along the lines of sight toward $\epsilon$ CMa and $\lambda$ Sco
presented in Appendix \ref{LISM_abundances}
(Table \ref{tab-ppm}, and also Table \ref{tab-Rgd}) the results in Figure 
\ref{figSF} can be
used to place constraints on the velocities of shocks which
accelerated and processed the local dust grains. As indicated by the
squares with error bars in Figure \ref{figSF}, the Si and Mg
abundances toward $\epsilon$ CMa indicate that the LIC dust grains have
been processed through a shock with a velocity V$\sim$110--220 km
s$^{-1}$. For this same line of sight the gas phase Fe abundances
indicate a shock of velocity V$\sim$90--100 km s$^{-1}$. The
latter velocity may be a poor indicator because the results of Jones
et al. (1996) are probably not valid for the shock processing of
dense, Fe-rich grains which would undergo enhanced sputtering destruction
(cf. the results for Fe grains in Jones et al. 1996).  
Indeed, if the data are to be believed then it appears that the
Fe and Mg in interstellar dust may be more resistant to erosion by
sputtering than Si.
For the $\lambda$ Sco line of sight the
indicated shock velocities are much smaller and are of the order of
V$\sim$0--70 km s$^{-1}$. 
The origin for this discrepancy may result partly because the warm cloud complex towards $\lambda$ Sco
is a blend of several components (\cite{welty96}), including both
the weakly depletely LIC and pockets of more depleted gas.

%\placefigure{figSF}

\section{Presolar and Interplanetary Dust Grains}\label{hoppe}
Our knowledge of the
elemental and isotopic composition, and grain history and sizes, for presolar dust grains found
in primitive meteorites,and for interplanetary dust particles (IDPs), is far more precise than our
understanding of the similar properties for interstellar dust grains.
We briefly discuss these
data sets in search of insights into the origin and properties of interstellar dust
grains.

\subsection{Presolar Grains}\label{presolar}
While most of the material that went into the making of the Solar System was thoroughly 
processed and mixed, thus losing isotopic heterogeneity and all memory of its origin, small 
quantities of refractory dust grains survived the events that led to the formation of the solar 
system and such grains have been found in primitive meteorites (\cite{1}; Zinner 1997; Ott 1993). They can be 
distinguished from other matter in the Solar System primarily on the basis of their 
anomalous isotopic compositions and they are believed to have formed in stellar outflows and 
in supernova ejecta. Presolar grains identified in primitive meteorites to date include 
diamonds, silicon carbide (SiC), graphite, silicon nitride (Si$_{3}$N$_{4}$), corundum (Al$_{2}$O$_{3}$), and 
spinel (MgAl$_{2}$O$_{4}$). Moreover, SiC and graphite grains carry small inclusions 
of Ti-, Mo-, and Zr-carbides (\cite{berna}).
Although the presolar grains are older than 4.6 billion years
they can provide useful insights into the properties of interstellar
grains. Cosmic-ray exposure ages of presolar SiC grains have been estimated
from the amount of cosmogenic $^{21}$Ne produced from cosmic-ray spallation of
Si. Ages of up to 130 (and possibly 2000) million years have been inferred
(Lewis et al. 1994), but uncertainties remain.

\subsection{Isotopic compositions and stellar sources}

\begin{figure}[ht]
\epsscale{0.6}
\plotone{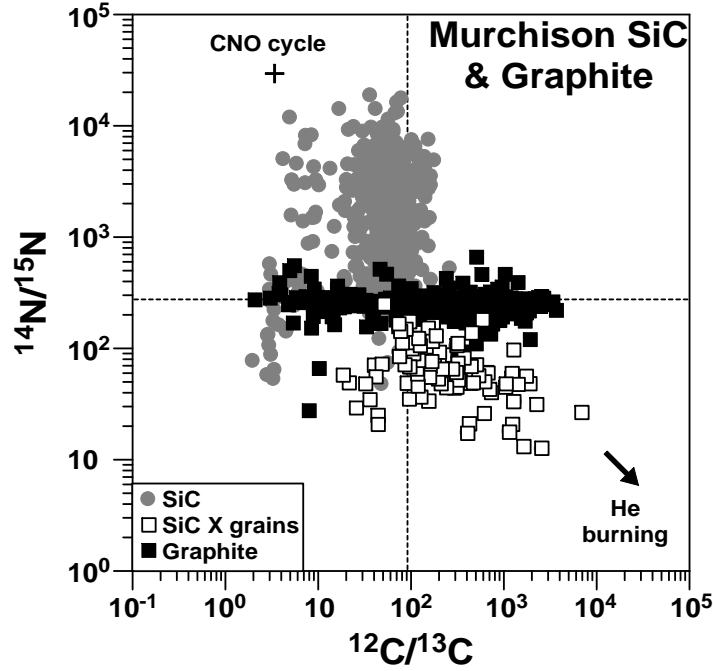} 
\caption{\it Carbon- and N-isotopic compositions
of SiC and graphite grains from the Murchison meteorite. The dashed
lines represent the Solar System isotopic compositions.  Theoretical
predictions for the CNO cycle and He burning are shown for
comparison. The grain data are from \protect\cite{5};
\protect\cite{7}; \protect\cite{8};
\protect\cite{9}.\label{hoppe.fig}}
\end{figure}

The physical and isotopic properties and likely stellar sources of the presolar grains from 
meteorites are summarized in Table \ref{hoppe.tab}. The diamonds are the most abundant but least 
understood presolar grains. One reason for this is their small average grain size of only 2 nm 
that does not permit the measurement of the isotopic compositions in single grains. The most striking 
isotopic signature of the diamonds is the presence of an anomalous Xe component, so-called Xe-
HL (\cite{3}). This component shows the signature of the p- and r-process of nucleosynthesis 
(photonuclear reactions and rapid n-capture, respectively), pointing to Type II supernovae as 
possible stellar source of at least a small fraction of the diamonds (\cite{4}). Most SiC grains 
("Mainstream", about 90\% of total) (\cite{5}) show imprints of the CNO cycle (cf. Fig. \ref{hoppe.fig}), some He 
burning, and the s-process (slow n-capture) and these grains are believed to have formed in 
the outflows of low-mass (1-3 M$_{sun}$) C-rich asymptotic giant branch (AGB) stars. The 
isotopic compositions of the rare SiC X grains (about 1\% of total SiC) 
(\cite{amazin}) and the isotopically 
related silicon nitride grains (\cite{7}) require matter from the interior of massive stars and Type II 
supernovae had been proposed as the most likely stellar sources of these grains.
Laboratory data indicate that SiC (in contrast to most silicates) dissociates
under bombardment by ionizing particle radiation, so that SiC may not survive as isolated grains in the ISM.
Graphite is isotopically distinct from SiC (Fig. \ref{hoppe.fig},
\cite{8,trav}). Most graphite grains show the signature of He 
burning and these grains appear to have formed in the outflows of massive stars, namely, 
Wolf-Rayet stars and, more likely, Type II supernovae. In addition, minor contributions from C-rich 
AGB stars and novae are evident. The isotopic compositions of most corundum and 
spinel grains (\cite{10,choi}) are well explained by models of red giant and AGB stars (e.g., Boothroyd et al.\ 1994) and 
the O-isotopic compositions agree quite well with those measured in the atmospheres of 
such stars (e.g., \cite{12}).

%\placetable{hoppe.tab}
\input{tabhoppe}

\subsection{Grain sizes}

Except for the diamonds, all types of presolar grains have sizes 
in the submicrometer- to micrometer-range (Table \ref{hoppe.tab}). 
These sizes are considerably larger than typical sizes of
0.005--0.25 $\mu$m for grains in diffuse clouds, which
provide the absorption and extinction of starlight in diffuse clouds.
However, presolar grains may contribute to the large end of the mass 
distribution determined from the {\it in situ} measurements (\S\ref{concl},\ref{massdis}), and radar micrometeor data (\cite{taylor96}).
While the measured sizes of SiC and graphite grains are 
considered representative for the true distributions in the meteorites, the measured sizes of 
the corundum and silicon nitride grains are probably biased and shifted towards larger grain 
sizes (\cite{10}) and are thus not representative for the true distribution within the meteorites. The 
size distribution of SiC has been studied in most detail. Variations among SiC from different 
meteorites are evident, pointing to the operation of a size sorting mechanism in the solar 
nebula. SiC from the Murchison CM2 meteorite tend to be larger than SiC from other 
meteorites. About 20\% of all Murchison SiC is in the $>$ 1 $\mu$m and only about 4\% in the $<$ 0.3 
$\mu$m fractions (Amari et al.\ 1994). SiC from the Indarch (EH4) meteorite is considered to be more 
representative for SiC in the protosolar nebula (Russell et al.\ 1997). Indarch SiC is finer grained and almost 
entirely composed of submicron grains with 60\% of the grains in the $<$ 0.3 $\mu$m fraction and 
only 4\% in the $>$ 1 $\mu$m fraction. 

\subsection{Presolar silicates and GEMS}\label{gems}

Presolar silicates have not been found so far in primitive meteorites. 
This is an outstanding puzzle, since interstellar dust grains with 
silicate mantles are required in diffuse material, including nearby ISM,
both by abundance arguments (\S\ref{licgrainsize}), and by observations
of polarization in nearby stars (Appendix \ref{polarization}).
Bradley (1994) has argued that the chemical anomalies found in GEMS 
(``glass with embedded metal and sulfides'') from 
interplanetary dust particles present a viable model for the structure and 
composition of interstellar silicate grains. In response to this 
hypothesis Martin (1995) presented a detailed analysis of the strengths 
and weaknesses in this hypothesis. In favour of GEMS being a good model 
for interstellar silicate grains are the following arguments:
1) the irradiated surfaces could be caused by cosmic rays and shocks in the ISM,
2) Fe, Mg and Si show differentiation consistent with interstellar depletions, 
3) they contain SPM inclusions that are required by interstellar polarization models, and
4) they are dirty enough silicates to explain the interstellar extinction and absorption. 
On the contrary they are not without problems, the most important of which are:
1) they contain sulphur, but sulphur is essentially undepleted in the ISM, and
2) meteoritic pre-solar grains show no evidence of interstellar radiation damage.

Nevertheless, the presolar nature of GEMS remains uncertain as long as no isotope data are 
available. Silicates are expected to be the dominant dust in circumstellar shells of red giant 
stars (\cite{15}). But these circumstellar dust grains may be smaller in size and have therefore remained undetected in the 
study of meteoritic matter or, alternatively, they may have been destroyed in the solar nebula 
or in the meteorite parent bodies. Analysis of cometary matter appears to be important for identifying presolar silicates
because comets are believed to contain the most primitive Solar System matter and a 
large fraction of the silicate grains contained in comets may be of presolar origin. This 
emphasizes the importance of the STARDUST mission which will allow us to analyze 
cometary matter in the laboratory.

%\placefigure{hoppe.fig}

\section{Discussion}\label{disc}

In this section we discuss the implications of the variations
in local gas-to-dust mass ratios found from the {\it in situ}
observations versus several astronomical sightlines.

\subsection{LISM:  $R_{\rm g/d}$ and Reference Abundance
\label{disc_g2d}}
The {\it in situ} data give an upstream direction of
interstellar grains which is close to the upstream direction of inflowing
neutral LIC gas, indicating that the interstellar cloud surrounding
the Solar System feeds these grains into the Solar System.
The comparison of LISM absorption line data (\S\ref{licgas2dust}) 
with {\it in situ} Ulysses and Galileo interstellar dust data
leads to several conclusions about grain properties.
The spacecraft observations show that dust is present in the LIC
but with larger typical size ($a_{\rm gr}$$\sim$0.5 $\mu$m for density
2.5 g cm$^{-3}$ and spherical grains) than typical MRN grains.
Since smaller grains are partially excluded by interactions with
heliospheric plasmas (\S\ref{exclusion}),
the $R_{\rm g/d}\sim 94^{+46}_{-38}$ value determined from grains measured
{\it in situ} will be an upper limit. 
This ratio depends on LIC properties, including
total density ($n$(H$^{o}$+H$^{+})$=0.3 cm$^{-3}$, \S\ref{LIC_model}).
Figure \ref{densdis} shows that small grains expected in the MRN distribution 
($m_{\rm gr}$$\sim$10$^{-15}$ gr, or $a_{\rm gr}\sim$0.05 $\mu$m for
spherical grains with density 2.5 g cm$^{-3}$), are in the range of the smallest masses of the grains detected 
by the {\it in situ} measurements, but the contribution of these small grains to the
mass-density is low compared to what we would expect from the larger grains
exhibiting an MRN distribution (Fig. \ref{densdis}). We partially
interpret the deficit of small grains
detected by the satellite observations, in comparison to the MRN
mass distribution, to the
interaction of these small grains with the solar wind magnetic field
(grains with $a_{\rm d}<0.1--0.2$ $\mu$m, \S\ref{hsfiltration}), and the interstellar
magnetic field in the heliopause region (grains with $a_{\rm d}<0.05$ $\mu$m, \S\ref{hpfiltration}). 
This indicates that although MRN particles must exist in the LIC,
they can not be measured in the inner heliosphere where the Ulysses
and Galileo measurements were made.
So, to calculate the actual gas-to-dust mass ratio for the LIC, we have to add the contribution of
the MRN particles, which were not detected {\it in situ}.
If we estimate
the gas-to-dust mass ratio of the LIC based on the
{\it in situ} data, and add in the missing mass for the
excluded MRN 
particles, then a value of $R_{\rm g/d}$=$56$ is found. 
The {\it in situ} detection is limited at large grain masses by
the bad statistics, since large grains are rare by number. If we
extend the mass distribution up to the mass of radar meteoroids
(\cite{taylor96}) the way we described it in Section \ref{totaldens},
the gas-to-dust mass ratio will drop to $31$.

In comparison, atoms missing from the gas phase in the LISM give 
gas-to-dust ratios of $R_{\rm g/d}$=427$^{+72}_{-207}$ for $\epsilon$ CMa,
and $R_{\rm g/d}$=137$^{+16}_{-40}$  for $\lambda$ Sco,
assuming solar reference abundances (\S\ref{licgas2dust}).
These two ratios are inconsistent with each other and both are larger than
the value from {\it in situ} detections.
The ratio is highly sensitive to the assumptions about the reference
abundance, and the assumption of B-star abundances increases the gas-to-dust
mass ratios determined from absorption lines data by factors 2--3, 
The uncertainties on the $R_{\rm g/d}$ values are large, due primarily to the
relatively small fraction of the material contained in the grains
(so that generally small fractional uncertainties in gas PPM values
yield large fractional uncertainties in dust PPM values). 
However, the $\lambda$ Sco and 23 Ori diffuse clouds, both with
H$^\circ$ column densities typically $\sim$19 dex or smaller,
strongly support large values for $R_{\rm g/d}$ for assumed B-star reference abundances.
The $R_{\rm g/d}$ values derived from the comparison of {\it in situ} and astronomical data yield 
somewhat better agreement for assumed solar, versus
B-star reference abundances, but in the best data for the cloud
feeding the grains into the Solar System, the data still is not adequate.
We conclude that the assumption that
all of the mass of LISM dust grains arises from the exchange of atoms with the
gas phase may be incorrect.
If B-star abundances are correct, we consider the possibility that part of the
total mass contained in interstellar dust and measured {\it in situ} may
not originate in the interstellar medium.
In the second case, we could in principle turn the problem around
and input known reference abundances to the ISM
$R_{\rm g/d}$ calculations, and compare the results with {\it in situ} data
to give a direct determination of the relative proportions of the
interstellar dust mass which may be attributed to direct stellar sources
versus condensation in the interstellar medium.  

\subsection{Shocks and Grain Models}\label{coremodel}
The LISM shows enhanced abundances of refractory elements in the
gas phase (\S\ref{lism_abs}) 
when compared to more distant cold clouds, which is
explained by the partial destruction of grains by shock fronts 
(\S\ref{SF}).  The destruction of grains by shocks explains the
larger value in nearby interstellar gas of the gas-to-dust mass
ratio,
$R_{\rm g/d}$, when compared to cold cloud values.  For instance,
the cold cloud complex towards $\zeta$ Oph has
$R_{\rm g/d}$=103$\pm^{+13}_{-29}$.  It also explains the good
correlation found between Fe and Mg column densities over several
order of magnitudes of material (Figure \ref{fe_mg_fig}). 
Even in the LIC over 80\% of Fe and 58\% of Mg is locked into dust grains,
causing small variations in dust mass to yield large variations in 
gas phase abundances.

When the composition of dust grains in the LIC towards $\epsilon$ CMa
is compared to models for grain destruction in shock fronts, 
shock velocities on the order of $\sim$100--200 km s$^{-1}$ are found.
In the LISM gas towards $\lambda$ Sco, with larger column densities,
lower shock velocities are found.  
A general model presents grains consisting of
a robust refractory core, coated with a mantle of more volatile material
(e. g. \cite{sofia94,ssa,ss,jones90,greenberg}).
The composition of the robust core has been modeled using observations of
selected clouds in halo star sightlines (\cite{ss}), 
indicating that the robust grain core consists of
mixed oxides and silicates, for either assumed solar
or B-star abundances.  If solar abundances are assumed, grain
mantle material appears to be silicates; if B-star abundances are assumed then
the core and mantle materials are not chemically distinct as would be
consistent with grain destruction by interstellar shocks.
This procedure is instructive, although recent discussions of interstellar
grains as conglomerates of fluffy or multi-sized grains suggest alternate
interpretations may be viable (e. g.\ \cite{dwek}).

%\placetable{tab-depl}

\subsection{Dynamic Separation of Gas and Grains}\label{dynsep}
The dust-to-gas ratios indicate the LIC may be inhomogeneous.
There are several processes by which interstellar grains couple to the
other components of the diffuse interstellar medium. The frictional
scale $l_{\rm drag}$ over which the dust couples to the gas is given
by the length it takes the grain to sweep-up its mass, $m_{\rm gr}$, in
form of interstellar gas (hydrogen, $m_{\rm H} = 1.67\cdot 10^{-24}\
{\rm g}$):
\begin{eqnarray}
l_{\rm drag} & = & \frac{m_{\rm gr}}{A_{\rm gr} (n_{\rm H} m_{\rm H} + n_{\rm He} m_{\rm He})}\label{eq-drag},
\end{eqnarray}
where $A_{\rm gr}$ is the cross-sectional area of the grain. This
frictional length scale is proportional to the size, $a_{\rm gr}$, of the grain
and exceeds the extent of the local cloud (a few pc) already for most
MRN-sized grains (e. g. $l_{\rm drag}$$\sim$14 pc for a grain with radius
$a_{\rm gr}$=0.1 $\mu$m). This implies that the gas and dust in the local
cloud are not in frictional equilibrium. 

However, there is a much shorter length scale due to the coupling of
the grains to the interstellar magnetic field. Interstellar grains in
the local environment are electrically charged to about $U \approx 1
~ {\rm V}$ surface potential due to the dominant photo-electric effect
from the background UV radiation field (\S\ref{graincharge}).  
With grain charge and gyroradius given by Eqs. \ref{eq-charge} and \ref{eq-gyro}, 
a typical relative speed, $v_{\rm rel}$ of several ${\rm km}\ {\rm s^{-1}}$ between the
grains and magnetic field of strength $B \approx 1.5\ \mu{\rm G}$, the gyroradius $l_{\rm gyr}$ is less than the size of
the local cloud even for micron-sized particles. Since the
interstellar magnetic field couples closely to the ionized component
of the interstellar medium, which in turn couples to the neutral component,
interstellar grains are indirectly coupled
to the gas. The effect is that gyrating dust particles have much
increased path lengths in the local cloud and hence gas drag can
thermalize the dust. However, if the relative speed between the magnetic
field and the dust is much higher, as it would be in a supernova shock crossing,
or the magnetic field is weaker, micron-sized grains
would no longer couple to the magnetic field over dimensions of the
local cloud. In these cases micron-sized and bigger particles would
decouple from small grains and from the gas cloud. The different apex
directions and much higher speeds of interstellar meteor particles
(where $10^{-9}$ to $10^{-7}\ {\rm g}$, {\cite{taylor96}) support such a scenario.

Big particles ($10^{-7}\ {\rm g}$, or $a_{\rm gr}$=5 $\mu$m at density 2.5 g cm$^{-3}$) can travel several $100\ {\rm pc}$
before they are captured by the magnetic field and hence by the gas,
i.e. big particles may be generated several $100\ {\rm pc}$ from the
place where they are found. It is evident that big particles couple
over much larger scales to the ambient diffuse interstellar medium
(gas and fields) than do small particles, therefore, they may be
compositionally unrelated to the gas and the small particles, with
independent origins.
Consequently the actual dust to gas mass ratio may vary locally, and hence, the cosmic
abundance ($1\% $ of the total mass in refractory dust) may be valid
only over large scales ($100$ to $1000\ {\rm pc}$).
This conclusion qualifies the total gas-to-dust mass ratio derived here
for the LISM but it does not qualify the relative comparison between the values derived from
the {\it in situ} versus interstellar gas observations, since
the grain size sampled {\it in situ} is dynamically bound to the gas being 
sampled by the interstellar absorption line data.

\subsection{Effects of Big Particles on the Collisional Evolution of
Interstellar Dust}\label{evolution}
The existence of large numbers of big particles (bigger than described by
the MRN distribution) in the diffuse interstellar medium has profound
consequences for evolution of interstellar material. Because of their
relatively small geometric cross section they do not show up in
optical observations.  The total mass of dust grains
not sampled by extinction measurements increases typically with grain radius,
i. e. $m_{\rm gr}/A_{\rm gr}$, so large grain particles  provide a significant
reservoir for the production of smaller particles by collisions.  This fact
may help solve the mystery of the rapid destruction times of MRN-sized
interstellar grains compared
to the transport and injection times of grains into the diffuse
interstellar medium.

Current models of dust processing in the ISM (Jones et al. 1996) indicate 
that grains with radii of order ten to a few thousand Angstroms 
(10$^{-3}$ to $\sim$0.3 $\mu$m)
are rapidly fragmented into smaller particles due to the effects of shattering in 
grain-grain collisions in interstellar shock waves. This disruption occurs on timescales 
of order tens of millions of years. Thus, the presence of grains with sizes 
ranging from a few thousand Angstroms to those of order one micron in 
meteorites (\S\ref{hoppe}), and in the detected interstellar dust in the Solar System, presents something 
of a challenge to current models of dust processing in interstellar shock waves.
However, these same models (Jones et al. 1997) have also shown that grains 
with radii much larger than one micron can survive fragmentation in shocks 
due to the underabundance of particles with radii large enough to cause their 
catastrophic disruption. 

In this context, the dust destruction level of 2/3 for the LIC implies
shock velocities greater than 200 km s$^{-1}$, and therefore the 
``large'' MRN grains may actually fare better because the postshock
compression is delayed and the large grain maximum velocities are reduced.  
In this case the thermal sputtering begins to destroy smaller grains.
This may provide a partial solution to both the large average grain
sizes observed {\it in situ} and the discrepant gas-to-dust mass ratios
problems.  If the LIC shock had velocity $>$200 km s$^{-1}$, the shock would
preferentially destroy small grains with respect to large grains by thermal sputtering.
In this case, the $R_{\rm g/d}$ value derived from {\it in situ}
measurements may require little correction for the mass of the
excluded grains, and therefore bringing the {\it in situ} determination closer
to the values measured for the LISM with interstellar absorption
lines.  This issue is important since interstellar spacecraft will
have to survive high-speed impacts with interstellar dust grains.

Two important parameters that determine the outcome of grain-grain
collisions are the mass ratio, $\Gamma$, of the target and projectile
grains and the relative velocity of the collision between the two. A
catastrophic collision has been defined as a collision in which more
than half of the larger (target) particle is shattered into smaller
fragments (Jones et al. 1996). For the MRN dust size distribution
(Equation 3) the mass range is about 10$^5$. However, for the most
disruptive collisions $\Gamma$ ranges from 10$^2$--10$^4$ for $\sim$
0.1 $\mu$m-sized target grains and $\sim$ 0.01 $\mu$m-sized projectile
grains (Jones et al. 1996). Many (but not all) of the grain-grain
collisions in a 100 km s$^{-1}$ shock result in catastrophic
shattering. The result is that a large fraction of the largest grains
are completely disrupted in collisions with smaller particles. This
leads to disruption timescales of about $5 \times 10^7$ years for
$\sim$ 0.1 $\mu$m grains and an accompanying steepening of the size
distribution (Jones et al. 1996). The disruption timescale is short
compared to the injection timescale of $2.5 \times 10^9$ years for the
formation and ejection of stardust into the diffuse ISM (Jones and
Tielens 1994). However, large grains can be re-formed in the dense
regions of the ISM during cloud collapse and there is strong evidence
for this grain growth in the denser ISM (e.g. Kim, Martin and Hendry
1994; Kim and Martin 1996).

In contrast, the particles observed by Galileo and Ulysses are as
large or much larger than the typical interstellar dust particles required to
explain the diffuse ISM extinction (e.g. Mathis 1990). It might be
expected that they therefore have longer lifetimes in the ISM due to
the underabundance of grains with radii sufficiently large to
catastrophically destroy them in supernova-generated shock
waves. However, this has yet to be demonstrated by detailed
modeling. Hence one might consider that the large particles observed
by Galileo and Ulysses could be considered as a conveyor belt that
carries dust mass from their circumstellar sources to the diffuse ISM,
where they become a reservoir for small particle formation by their
fragmentation in grain-grain collisions in shock waves. Their ultimate
demise would signal the end of the conveyor belt. Since big particles
do not couple as closely to the gas as small grains (\S\ref{dynsep}) they
may form a background large grain dust distribution. Thus these
interesting scenarios will have very important implications for the
inhomogeneity of the ISM and the competition between large grain
survival and dust re-cycling in the ISM. These complex processes would
be worth investigating in greater detail.

\section{Conclusions }\label{concl}
In this paper we have produced five values, summarized in Table \ref{tab-Rgd}, 
for the gas-to-dust mass ratio for the local interstellar cloud.  The first value 
is
determined from {\it in situ} observations of interstellar dust made by the
Ulysses and Galileo spacecraft, and gives $R_{\rm g/d}$=94$^{+46}_{-38}$
for the interstellar dust grains which are able to penetrate the
heliosphere (typical grains with radii greater than 0.2 $\mu$m).  A set 
of values is determined for nearby interstellar gas based on interstellar 
absorption line measurements towards two stars, with the assumption that
a standard reference abundance exists which describes the relative 
elemental
abundances for the interstellar material, gas plus dust combined.  This 
assumption is equivalent to assuming that all of the dust mass arises from
the exchange of mass with interstellar gas.  When the reference abundance is 
assumed to be that of B-stars, values for $R_{\rm g/d}$ that are 4--5 times 
larger than the {\it in situ} value are found
for the clouds considered towards $\epsilon$ CMa and $\lambda$ Sco (Table \ref{tab-Rgd}).   If 
the
reference abundance is instead assumed to be solar, then the value for 
$R_{\rm g/d}$ found towards $\lambda$ Sco are comparable to the {\it in 
situ}
value, considering uncertainties.  However, the cloud best
representative of the cloud surrounding the Solar System is the LIC cloud 
towards $\epsilon$ CMa, and $R_{\rm g/d}$=427$^{+72}_{-207}$ found here 
is still a factor of three greater than the {\it in situ} value.  
Therefore, we conclude that although solar abundances yield the best match 
between astronomical and {\it in situ} determinations of $R_{\rm g/d}$, 
regardless of the selected reference abundance it is likely
that the interstellar dust grain population detected {\it in situ} contains 
grains which have not been formed exclusively by mass exchange with the LIC gas.  This conclusion is 
entirely consistent with models of interstellar dust grains which invoke
circumstellar and supernova sources for the interstellar dust grain 
population,
although models of such sources are generally inadequate to explain 
observed grain
abundances (e. g. \cite{dwek98}).   

Both the {\it in situ} and extinction determinations of $R_{\rm g/d}$
consider grains from all sources, whereas the $R_{\rm g/d}$
calculated from ``missing'' interstellar gas does not include grain
mass derived from stellar sources or any dust grain large
enough to be inefficiently destroyed by shock fronts.
Presolar grain data show us that
stellar sources contribute over a large size range (\S\ref{hoppe}),
and therefore do not violate the {\it in situ} data.
Based on these discussions, some of the dust mass observed {\it in situ}
must arise from an additional source which does not exchange mass
with gas in the interstellar cloud surrounding the Solar System.

One possibility is that interstellar dust grains are seeded by refractory
presolar-type grains from cool stars and supernova (\S\ref{hoppe}).  If this were true, then
we could in principle turn the problem around
and input known reference abundances to the
$R_{\rm g/d}$ calculations, and compare the results with {\it in situ} data
to give a direct determination of the relative proportions of the
interstellar dust mass which may be attributed to direct stellar sources
versus condensation in the interstellar medium. However, the LIC gas has 
been shocked by high velocity shocks (\S\ref{SF}) so that a pre-solar type 
grain population is also likely to have been shocked.
Very little mantle material remains on the LIC dust grains because
of destruction of these grains in shock fronts.
A good correlation is found between 
Mg$^+$ and Fe$^+$ in nearby interstellar gas, suggesting that
the mineral composition of the parent dust grains dominates the gas-phase
abundances of these elements.  
In contrast, the poor correlation between Fe$^+$ and H$^\circ$ 
suggests either inhomogeneous ionization or poor dust-gas mixing on 
small spatial scales.

A second possibility is to look at the shock process itself.
Large grains are inefficiently destroyed by shocks (\S\ref{evolution}).
Therefore pockets of large grains may remain from the giant molecular clouds
from which the subgroups of the Scorpius-Centaurus Association formed,
with the grains captured in, and displaced towards the Sun, by the expanding 
superbubble shell surrounding this
association (\cite{fr95}).  This process may contribute larger grains which have not
exchanged mass recently with the LISM, since large grains are less
likely to be thermalized in shock fronts (\S\ref{evolution}).

Once we are forced to invoke a distinct population of interstellar dust grains 
that does not interchange mass with interstellar gas, then we can not 
conclude that the
better $R_{\rm g/d}$ ratios found for the LISM data and assumed solar 
abundances indicate that
solar abundances are to be preferred over B-star abundances.  On the other
hand, if the $\lambda$ Sco cloud complex were to be found to be
relatively homogeneous, we would propose that the LISM cloud complex 
towards $\lambda$ Sco, rather than the LIC seen towards $\epsilon$ CMa, is
the cloud feeding dust into the Solar System.  This is similar to saying
that the Sun is immersed in the same interstellar cloud as $\alpha$ Cen,
which does not appear to be the case.

Since we will conclude that the enhanced refractory abundances in the
nearby interstellar gas indicate that the dust grains in the LIC have
been shocked,
it is of interest to search for correlations between the upwind direction
of the LISW and external galactic kinetic phenomena.  Frisch (1981)
concluded that the Sun is located in the expanding superbubble
shell from the Loop I supernova, based on the velocity of gas
and enhanced refractory abundances in the LISM.  When solar
motion is removed from the LISW heliocentric flow vector,
the local interstellar material is seen to flow from a direction
corresponding to $l$=315$^\circ$ to 325$^\circ$, $b$=0$^\circ$ to --3$^\circ$
(depending on the assumed solar motion; Frisch 1995; Table \ref{licprop}).  
This direction is close to the
direction of the center of the Loop I 21 cm H$^\circ$ shell
located near $l$=320$^\circ$, $b$=+5$^\circ$ (Heiles 1998).  Based on the
LIC dynamics and shock models (\S\ref{SF}), the
hypothesis that the Sun is embedded in a fragment of the expanding
Loop I superbubble shell thus remains a viable hypothesis.
 
The results presented in this paper indicate that
{\it in situ} measurements of the interstellar dust population in the
heliosheath and interstellar medium would yield a high scientific return.
When we have compared the gas-to-dust mass ratio found from the LIC
cloud with the value expected from MRN grain populations only,
we find 98\% of the MRN-sized particles are missing from the 
population observed by Ulysses and Galileo, evidently because of exclusion
from the heliosphere due to interactions with heliospheric plasmas.
The MRN mass spectrum is selected
as an example of an empirically derived interstellar dust grain distribution.
These MRN-sized particles are the interstellar grain population most likely
to be exchanging mass with the interstellar medium.
Measurements of the interstellar dust grain population outside of the
heliosphere and in the heliosheath regions would capture the
mass distribution of these interstellar particles excluded from the
heliosphere.  Comparisons of this distribution with the grain
properties inferred from absorption line measurements of nearby
stars would thus provide a valuable insight into the separate
histories of large versus small interstellar dust grains, and into 
the sources of large interstellar dust grains which have not exchanged mass
with the diffuse cloud surrounding the Solar System.
If separate origins and history of large versus small interstellar
grains are ascertained, and a grain population which does not exchange
mass with the interstellar medium can be identified, this information 
would aid in understanding the chemical evolution of the Milky Way galaxy,
external galaxies, and QSO absorption line systems.

%%ACKNOWLEDGEMENTS
%\input{acknowledgements}
\acknowledgements

The authors of this paper would like to thank the International
Space Sciences Institute (ISSI) in Bern, Switzerland, for hosting workshops
in January and October 1997 on the topic of ``Interstellar Dust in the Solar System''.
In particular, 15 of the co-authors would like to thank the 16th, Professor
Johannes Geiss, for hosting this series of workshopos.
The results presented in this paper are based on the
scientific discussions at these workshops.
We would like to thank NASA grant NAG5-6625 for assisting US participants 
in travel support for the workshops.  
The authors gratefully acknowledge support for their research.
GPZ was supported in part by a NASA grant NAG5-6469, JPL contract 959167
and a NASA Delaware Space Grant College award NGT5-40024. 
PCF has been supported by NASA grants NAG5-6405 and NAG5-6405. 
This research has made use of the Simbad database, operated at CDS, Strasbourg, France.

%%APPENDICES
\clearpage
\appendix
\section{Adopted LIC Model}\label{ap_LIC}
The properties of the interstellar cloud surrounding the Solar System
are listed in Table \ref{licprop}.  The basis for the choice of these
values is given in this section.  The physical properties of the
cloud surrounding the Solar System can be derived from observations
of interstellar matter within the Solar System (backscattered H$^\circ$ 
and He$^\circ$ radiations, direct detection of He$^{\circ}$, and the pickup
ion and anomalous cosmic ray particle populations), from observations
of ultraviolet and optical absorption lines in nearby stars, and from
extreme ultraviolet observations of nearby white dwarf stars.

Ulysses observations of He$^{\circ}$ within the Solar System give direct
measurements of interstellar neutral He$^{\rm \circ}$ in the cloud surrounding the
Solar System, since He$^{\rm \circ}$ is little affected by the traversal
of the outer heliopause and heliosphere regions (e. g. \cite{rucfahr89}).
The interstellar He$^{\circ}$ density found from these observations
is $n$(He$^{\circ}$)=0.016 cm$^{-3}$, and the downstream direction and 
relative Sun-cloud velocity were determined to be
$\lambda$=74.7$^{\circ}$$\pm$1.3${^\circ}$, $\beta$=--4.6$^{\circ}$$\pm$0.7$^{\circ}$
(ecliptic coordinates), and
V=24.6$\pm$1.1 km s$^{-1}$ (\cite{witte96}; Witte private communication).  
These values are confirmed by
EUVE observations of the 584 ${\rm \AA}$ backscattered glow
(downwind direction $\lambda$=76.0$^{\circ}$$\pm$0.4${^\circ}$, $\beta$=--5.4$^{\circ}$$\pm$0.6$^{\circ}$,
V=26.4$\pm$1.5 km s$^{-1}$, \cite{flynn}).  These values are converted
into heliocentric and LSR velocities, in the galactic coordinate
system, in Table \ref{licprop}.
The upwind direction in the LSR ($l$$\sim$315$^\circ$, $b$$\sim$--3$^\circ$)
is close to the center of the Loop I 21 cm superbubble shell
(at l=320$^\circ$, b=+5$^\circ$, Heiles 1998).  
Based on these dynamics, combined with the history of 
shocked grains in the LIC (Section \ref{SF}), the conclusion that 
the Sun is embedded in a fragment of the expanding
Loop I superbubble shell (Frisch 1981, 1995) remains a viable hypothesis.

Observations of nearby white dwarf stars in the extreme ultraviolet 
give a mean value
\begin{eqnarray}
<\frac{\rm N(H^{o})}{\rm N(He^{o})}> = 14.7
\end{eqnarray}
where $N$(X) is the column density (cm$^{-2}$) of element X
(\cite{fr95,dupuis,vallerga}).
For the stars with the two lowest foreground interstellar column
densities, GD 71 and HZ 43, the ratios are 12.1 and 15.8, respectively.
Combining these values, $n$(H$^{\circ}$)=0.19--0.25 cm$^{-3}$ 
is inferred for the H$^{\rm o}$ spatial density in the LIC cloud,
and we adopt the value $n$(H$^{\circ}$)=0.2 cm$^{-3}$.
However, the LIC cloud is ionized, with the relative ionizations of H and He
changing as a function of cloud depth.  
Hydrogen ionization increases towards the cloud edge with respect to
helium ionization because of cloud optical depth effects. Therefore the ratio
$N$(H$^{\circ}$)/$N$(He$^{\circ}$) at the solar location should in
principle be larger than the sightline
averaged values if surrounding interstellar material is homogeneous,
which in turn allows $n$(H$^{\circ}$) to be larger than this adopted value.
Although inhomogeneities are clearly present (\S\ref{LICdust}, \ref{disc}), we do not consider the possibility that the inhomogeneities would
yield higher neutral densities for the LIC.
Radiative transfer models of the
LIC give an electron (or proton) density of $n$(e$^{\rm -}$)$\sim$0.11 cm$^{-3}$
at the solar location (Slavin and Frisch 1997), a value in agreement with LIC values
$n$(e$^{\rm -}$)=0.09 cm$^{-3}$ (+0.23, -0.07) found towards $\epsilon$ CMa (Gry and Dupin 1995) 
and $n$(e$^{\rm -}$)=0.11 (+0.12, -0.06) cm$^{-3}$ found towards $\alpha$ Aur (Wood and Linsky 1997), and $n$(e$^{\rm -}$)=0.11 cm$^{-3}$ towards
$\eta$ UMa (\cite{fr98uma}).
This gives values $n$(H$^{o}$)=0.21 cm$^{-3}$ and $n$(e$^{\rm -}$)=0.11 cm$^{-3}$ 
for the LIC cloud, yielding a total LIC density of 0.32 cm$^{-3}$.
The LIC temperature determined from recent observations of He$^{\circ}$ 
backscattered radiation in the Solar System is 6,900$\pm600$ K 
(Flynn et al. 1997), and this value is consistent with interstellar absorption line data
in nearby stars (e. g. \cite{lalstruc}; 
\cite{gry96}; \cite{linstars}).  Observations of the neutral He flux by Ulysses give
T=6,100$\pm$300 K (\cite{witte96}; Witte private communication).  

The LIC magnetic field strength has not been directly determined.
We adopt a magnetic field strength B=1.5--2.0 $\mu$G because
this value is consistent with the ordered
component inferred from global pulsar data (\cite{fr90}), 
comparisons between models and observations of hydrogen-wall L$\alpha$ absorption 
towards $\alpha$ Cen (\cite{gayley}), pick-up ion data constraints 
(\cite{gloeck97}), and magnetohydrodynamic models 
of the heliosphere (\cite{linde}, \S \ref{graincharge}).

%\placetable{tab-ppm}
\input{table_ppm}
\clearpage

The physical conditions in the LIC are equivalent to those in the
warm partially ionized medium, or intercloud medium, in the
McKee and Ostriker (1977) three-phase model of the
ISM. Thus, the Jones et al. (1994, 1996) results on grain destruction 
by interstellar shock fronts are directly
applicable to the LIC conditions, and conversely the nature of the
dust and gas in the LIC provides a rigorous test of these dust
processing models.

\section{Upstream Direction from {\it In Situ} Detections}\label{direction}
The kinematic relationship between dust and gas in the LISW can be
revealed by looking at the velocity dispersion of the grains measured
by Galileo and Ulysses.   
The upstream direction and relative Sun-cloud
velocity have been found from observations of interstellar
He$^\circ$ within the Solar System, both by direct observations
of the neutral atoms with the Ulysses neutral gas experiment
(\cite{witte96}) and by observations of the backscattered
He$^\circ$ 584 ${\rm \AA}$ radiation (e. g., \cite{flynn}). 
These directions are given in Appendix \ref{ap_LIC}.

The interstellar influx direction was found from {\it in situ} data
collected with the Ulysses spacecraft, and compared
with the neutral gas and backscatter results.  Since the Ulysses detector
has a wide field-of-view ($\pm 70^\circ$), the impact
direction for individual events can not be directly determined, but
is found rather from the statistical properties of the ensemble.
To extrapolate from the impact direction to the upstream direction
outside the heliosphere, we assume the simplest case that the
particles have straight trajectories from the interstellar medium to
the point where they are detected. Doing this, we neglect the grain dynamics
within the Solar System. This assumption leads to an error which is
smaller than the statistical error in the determination of the
direction. 

We apply a $\chi^2$-fit to the histogram of detected rotation angles
with the ecliptical longitude and latitude of the upstream direction
as the parameter set. Since the line of sight of the Ulysses detector
was scanning mainly latitudes (for trajectory and geometry see
\cite{gruen93}), the latitude of the upstream direction is determined
with higher accuracy. Figure \ref{chi2} shows a contour plot of
$\chi^2$ in the two-dimensional parameter space of ecliptic coordinates.

%\placefigure{chi2}
\begin{figure}[ht]
%\figurenum{chi2}
\epsscale{0.8}
\plotone{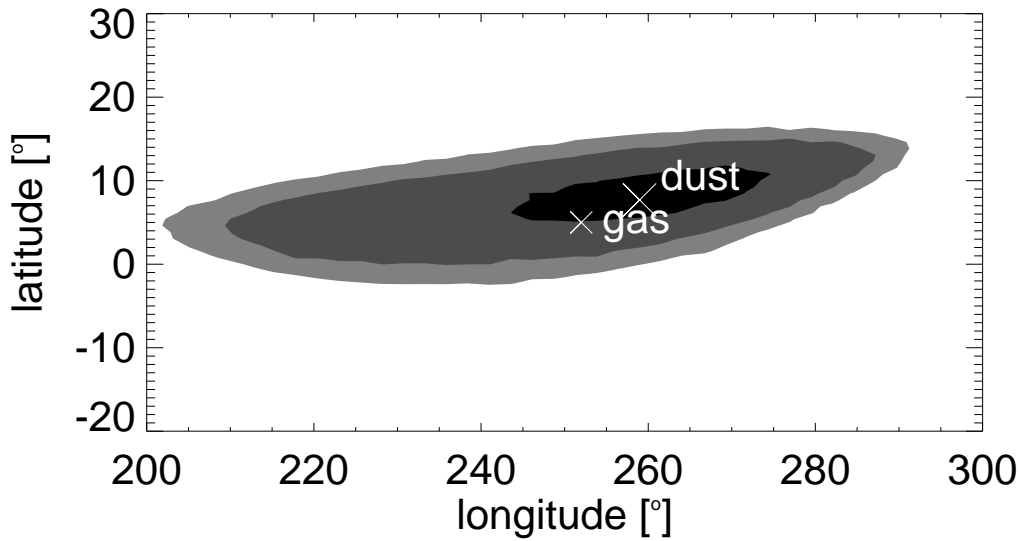}
\caption{\it Contour plot, in ecliptic coordinates, of the upstream
direction of interstellar dust grains detected by the Ulysses and
Galileo satellites.  Contour levels $1\sigma$ (black), $2\sigma$ (mid
gray), and $3\sigma$ levels (light gray) of $\chi^2$ are shown.  The
parameter space is defined by ecliptic longitude ($\lambda$) and
ecliptic latitude ($\beta$). The upstream direction of He$^{\rm o}$
$(\lambda=254^\circ,\beta=+5.6^\circ)$ is shown by the ``gas'' label
(Witte et al.\ 1993).  The ``dust'' label shows the position of
minimal $\chi^2$ for the dust upstream direction
$(\lambda=259^\circ,~\beta=+8^\circ)$.\label{chi2}}
\end{figure}

The contours shown in Figure \ref{chi2} have a smaller extent in the
direction of latitude than in direction of longitude and show that the
latitude of the upstream direction is determined with higher
accuracy. We give the upstream direction of interstellar particles as: ecliptic longitude 
$\lambda = 259^{\rm o} \pm 20^{\rm o}$, ecliptic latitude 
$= 8^{\rm o} \pm 10^{\rm o}$ ($1\sigma$-level uncertainties). The upstream direction of the
interstellar gas lies inside the $1\sigma$-range.

\section{Polarization by LISM Dust}\label{polarization}
Observations of the weak polarization ($\geq$0.05\%) of the light of nearby (d$<$35 pc) stars,
caused by non-spherical dust grains aligned by an interstellar magnetic field in the 
galactic interval $l$=340$^{o}$$\pm$70$^{o}$, $b$=0$^{o}$$\pm$40$^{o}$ (\cite{tinb}),
\footnote{This spatial interval corresponds to the
upwind direction of the ``local fluff'' cloud complex in the local
standard of rest velocity frame (\cite{fr95}).}  
are consistent with
LISM dust grains in the size range measured by Ulysses and Galileo. 
The standard grain diagnostic for the size distribution of polarizing
grains is the wavelength of maximum polarization (\cite{serkowski}), 
which is not known for the LISM grains, therefore
we can not make detailed predictions based on the polarization data.
The LIC apparently extends only a fraction of a parsec from the 
Sun in this region (\cite{lalstruc}).  However, the polarization strength is independent
of star distance indicating that both the dust grains and magnetic field
extend to the solar vicinity (\cite{fr90}).  
We will assume that the polarizing grains in the LIC have
properties consistent with the 
global properties for the polarizing grain population in diffuse clouds.
Possibly the polarizing grains are located in the more depleted pockets of gas
implied by the gas-to-dust mass ratio of $\lambda$ Sco.

We obtain a minimum amount of information from the polarization data.
Spatial variations in the wavelength of maximum polarization ($\lambda_{max}$)
between sightlines with similar extinctions indicate that the polarizing grains
are a subset of the dust grain population, with the remaining grains 
only contributing to the extinction
(Mathis 1986), therefore we adopt the characteristics of the
polarizing grains, rather than the extinguishing grains, as a
model for the LIC dust.  
This gives a low mass cutoff in the size distribution.
Mathis (1986) has calculated the size distribution for the
polarizing grain populations by fitting $\lambda_{max}$,
where the criteria for the size of the polarizing grain was that it
be large enough to include a superparametric (SPM) cluster (which
enhances the imaginary part of the magnetic susceptibility).
These relatively large polarizing grains are consistent with 
a deficiency of small grains seen in 
an MRN size distribution likely to have SPM inclusions was found to fit the observed range
of $\lambda_{max}$=0.35--0.72 $\mu$m, providing that the small end of the 
size distribution was cut off at a grain radius of $a_{\rm gr}$=0.03--0.16 $\mu$m.  
The more recent model of polarization grains of Kim and Martin
(1995) also requires relatively large grains to reproduce infrared
data.
Thus, if models for the global properties of polarizing dust grains applies to
the cloud surrounding the Solar System, 
then the properties of grains in nearby
interstellar material in the upwind direction are consistent with the Ulysses and Galileo
observations (section \ref{massdis}).  Recently, this model has been supported by Goodman and Whittet (1995), who argue that
SPM inclusions in GEMS (\S\ref{gems}) have the correct sizes ($\sim$0.1$\mu$m) to be
the polarizing grain component.

\section{LISM Abundances}\label{LISM_abundances}
\subsection{LISM Abundances and Depletions}\label{ap_depl}
The abundances and depletions of Fe and Mg towards nearby stars
are listed in Table \ref{tab-femg}.  These values are based on
the assumption of solar abundances.  The depletion of an element X with respect to H is defined as $\delta_{\rm H}$=$N$(X)/$N$(H)$_{observed}$ -- $N$(X)/$N$(H)$_{\rm\rm  ref}$, where $N$(X)/$N$(H)$_{\rm ref}$ is the reference abundance of the species X (using the standard
definition). 
The depletions are seen to
vary by $\sim$0.7 dex between sightlines in the LISM if solar reference abundances are used
and the comparison abundance is H$^{\circ}$.
For example, solar abundances give $\delta_{\rm H}$(Mg)=--0.44,
$\delta_{\rm H}$(Si)=--0.25, and $\delta_{\rm H}$(Fe)=--0.72.
The O and N in the LIC cloud towards $\epsilon$ CMa indicate
O and N are weakly depleted.
These values are plotted in Figures 
\ref{fe_mg_fig} and \ref{fe_hI_fig}.

%\placetable{tab-femg}
\input{table_star_abund}

The gas-to-dust mass ratios given in Table \ref{tab-Rgd} are based
on the column densities presented in Table \ref{tab-ppm}.

%\placetable{tab-ppm}
\subsection{ISM towards $\epsilon$ CMa and $\lambda$ Sco}\label{LICcolumns}
Table \ref{tab-ppm} gives absorption line data for the LIC component
at --17 km s$^{-1}$ towards $\epsilon$ CMa (from \cite{dupin98,gry95,gry96,gry98}), and the LISM gas towards
$\lambda$ Sco (from \cite{york83}).  Note that an order of magnitude more material is found towards
$\lambda$ Sco than $\epsilon$ CM (Table \ref{tab-Rgd}).
The $\lambda$ Sco data are {\it Copernicus} data, they may sample a cluster of
smaller ``clouds'' at closely spaced velocities.
Gas-phase and dust abundances, based on both solar and B-star reference
abundances for both $\epsilon$ CMa and $\lambda$ Sco are
presented in Table \ref{tab-ppm}.  Respective PPM values are
calculated for these stars, where the column densities labeled
``solar'' and ``B-star'' in Table \ref{tab-Rgd} are used to calculate
the PPM values labeled ``solar'' and ``B-star'', respectively.
Since oxygen and nitrogen ionization are closely tied to hydrogen ionization
by charge exchange, PPM values for O$^{\rm o}$ and N$^{\rm o}$ are calculated
relative to $N$(H$^{\rm o}$) while the PPM values listed for ions
Si$^{+}$, Mg$^{+}$, Fe$^{+}$, S$^{+}$, are calculated relative to 
$N$(H$^{\rm o}$)$+N$(H$^{\rm +}$).  Helium is not included
in the PPM estimates, but represents less than a 10\% correction.
These enhanced abundances of refractories exhibited by the LIC are discussed in
\ref{lism_abs}.
Comparing these values to the range of Mg and Si depletions found by
Fitzpatrick (1997), it is seen that the cloud around the Solar System is a weakly depleted
cloud, in agreement with Frisch (1981) and GD conclusions.  

The results obtained from the component 2 neutral cloud observed towards
$\lambda$ Sco by the {\it Copernicus}
satellite (\cite{york83}) are also listed in Tables \ref{tab-Rgd} and \ref{tab-ppm}.
This component, which contains all of the
neutral gas in the sightline,
coincides in velocity with the LISM gas in the
galactic center hemisphere (\cite{fy91}).  
In this case, the assumption
of B-star reference abundances is in clear disagreement with the
gas-to-dust mass ratio derived from the {\it in situ} data.

%\input{end}

%BIBLIOGRAPHY
\input{bibliography}

%FIGURE CAPTIONS
%\input{figure_captions}

%TABLES
%\clearpage
%\input{tabLIC}
%\clearpage
%\input{table_g2d}
%\clearpage
%\input{tabdepl}
%\clearpage
%\input{tabhoppe}
%\clearpage
%\input{table_ppm}
%\clearpage
%\input{table_star_abund}
%\clearpage

%FIGURES
%\clearpage
%\plottwo{massdis_g.ps}{massdis_u.ps}
%\clearpage
%\plotone{densdis.ps}
%\clearpage
%\plotone{helio_Bfield.eps}
%\clearpage
%\plotone{helio_gyro.eps}
%\clearpage
%\plotone{fig_mg.eps}
%\clearpage
%\plotone{fe_hI.eps}
%\clearpage
%\plotone{figSF.eps}
%\clearpage
%\plotone{fig-hoppe.ps}
%\clearpage
%\clearpage
%\plotone{chi2.eps}
\end{document}

%% file: tabLIC.tex
\begin{table}
\small
\begin{center}
\caption{Properties of Interstellar Cloud Surrounding Solar System \label{licprop}}
\begin{tabular}{lccc} 
\tableline
\tableline
Item&Adopted Values&&Notes\\
\tableline
$n$(He$^{o}$)&0.015 cm$^{-3}$&&1\\
$N$(H$^{o}$)/$N$(He$^{o}$)&14.7&&2, 3, 4\\
$n$(H$^{o}$+H$^{+}$))/($n$(He$^{o}$+He$^{+}$)&10&&10\\
$n$(H$^{o})$&0.22 cm$^{-3}$&&Inferred, 15\\
$n$(H$^{+}$)&0.10 cm$^{-3}$&&5, 6, 7, 12, 15\\
\multicolumn{4}{l}{Downstream direction in solar rest frame}\\
Ecliptic coordinates&$\lambda$=74.7$^{\circ}$$\pm$1.3${^\circ}$, $\beta$=--4.6$^{\circ}$$\pm$0.7$^{\circ}$&&1\\
&V=24.6$\pm$1.1 km s$^{-1}$&&\\
Upstream directions:&&&\\
Solar rest frame, galactic coordinates&$l$=2.7$^{\circ}$, $b$=+15.6$^{\circ}$ &&\\
&V=--24.6$\pm$1.1 km s$^{-1}$&&\\
LSR rest frame, galactic coordinates:&$l$=315$^{\circ}$, $b$=--3$^{\circ}$& &14\\
&V=--19.4 km s$^{-1}$&&\\
Temperature&6,900 K&&11\\
Magnetic Field&1.5--6 $\mu$G&&9, 13, Appendix \ref{ap_LIC}\\
\tableline
\end{tabular}
\end{center}
Notes:  
1.  \cite{witte96}, and Witte (private communication).
2.  \cite{dupuis}. 
3.  \cite{fr95}. 
4.  \cite{vallerga}.
5.  \cite{slafr}.
6.  \cite{gry96}. 
7.  \cite{woodlin}.
8.  Section \ref{intro}.
9.  \cite{fr90} (estimated value).
10.  \cite{ss}.
11.  \cite{flynn}.
12.   \cite{lalfer}.
13.  Linde (private communication).
14.  This value is based on subtraction of the ``standard'' solar motion of 
19.5 km s$^{-1}$ towards the direction $l$=56$^{\circ}$, $b$=+23$^{\circ}$ 
from the observed LISW heliocentric velocity vector.  
If the ``best'' solar motion of 
16.5 km s$^{-1}$ towards $l$=53$^{\circ}$, $b$=+25$^{\circ}$ had been subtracted instead, the
LSR inflow direction of the local interstellar wind would be
V=--18.2 km s$^{-1}$ from 
$l$=324$^{\circ}$, $b$=--1$^{\circ}$ (e. g. Frisch 1995).
15.  \cite{puy}
\end{table}

%% file: table_g2d.tex
\begin{deluxetable}{cccccc}
\footnotesize
\tablewidth{0pt}
\tablecaption{Gas-to-Dust Mass Ratios, $R_{\rm g/d}$\label{tab-Rgd}}
\tablehead{
\colhead{}&\colhead{{\it In situ}}&\colhead{$\epsilon$ CMa\tablenotemark{1}}&\colhead{$\epsilon$ CMa\tablenotemark{1}}&\colhead{$\lambda$ Sco\tablenotemark{2}}&\colhead{$\lambda$ Sco\tablenotemark{2}}\nl
\colhead{Ref. Ab.}&U/G\tablenotemark{3}&\colhead{Solar}&\colhead{B-star}&\colhead{Solar}&\colhead{B-star}
}
\startdata
log $N({\rm H^o})$&&17.30&17.51$\pm$0.07&19.23$\pm$0.05&19.23$\pm$0.05 \nl
log $N({\rm H^+})$&&16.95\tablenotemark{4}&17.15$\pm$0.20&& \nl
log $N({\rm H^o + H^+})$&&17.46&17.66$\pm$0.21&19.23$\pm$0.05&19.23$\pm$0.05 \nl
$R_{\rm g/d}$\tablenotemark{5}&$94^{+46}_{-38}$&427$^{+72}_{-207}$&551$^{+61}_{-251}$&137$^{+16}_{-40}$&406$^{+58}_{-243}$ \nl
\tablenotetext{1}{The column densities are from \cite{dupin98} and \cite{gry96};
the uncertainties are comparable to 2$\sigma$--3$\sigma$ (Gry, private communication).}
\tablenotetext{2}{\cite{york83}}
\tablenotetext{3}{{\it In situ} values are determined from Ulysses
and Galileo spacecraft data (\S\ref{totaldens}).}
\tablenotetext{4}{In this case, the ionized column density is estimated to be
$log N$(H$^{\rm +}$)=16.95 cm$^{-2}$, based on
$n$(e$^{\rm -}$)=$n$(p$^{\rm +}$)=
0.1 cm$^{-3}$ and $n$(H$^{\circ}$)=0.22 cm$^{-3}$.
These values for $R_{\rm g/d}$ are calculated using the column densities in Table \ref{tab-ppm}.
}
\tablenotetext{5}{$R_{\rm g/d}$ is defined in eq. 9.  The gas-to-dust mass ratio for $\lambda$ Sco
includes an estimate for Mg$^{+}$ column densities based on
the observed Fe$^{+}$ column density and the correlation
in Fig. \ref{fe_mg_fig}.
The uncertainties on $R_{\rm g/d}$ are discussed in \S\ref{licgas2dust}.
For comparison, the diffuse WLV component towards 23 Ori (Welty et al. 1999
data, see \S\ref{licgas2dust})
gives values $R_{\rm g/d}$=127$^{+13}_{-12}$ and $399^{+126}_{-120}$
for assumed solar and B-star abundances, respectively.
}
\enddata
\end{deluxetable}

%% file: tabdepl.tex
\begin{table}
\begin{center}
\caption{PPM Abundances for Core-Mantle Dust Grains $^{a}$\label{tab-depl}}

\begin{tabular}{lrrrrrrrrr} 

\tableline
Element&&Solar&Total&Core$^{b}$&Mantle&B-Star&Total&Core$^{b}$&Mantle\\
&&&Dust&&&&Dust&&\\ 
\tableline
&&&\\
{\bf LIC}&&&\\
Mg&&38&28&27&1&25&18&12&6\\
Si&&36&11&16&0&18.6&2.7&2.9&0\\
Fe&&32&27&25&2&27&24&13&11\\ 
{\bf $\zeta$ Oph Cold Cloud}$^{c}$&&&\\
Mg&&38&37&27&10&25&23&12&11\\
Si&&36&34&16&18&18.6&21&2.9&18\\
Fe&&32&32&25&7&27&20&13&7\\ 
\end{tabular}
\end{center}
$^{a}$  These data are based on the LIC component towards $\epsilon$
CMa (Table \ref{tab-ppm}).  
The units are PPM, which represents the number of trace elements per 10$^{6}$ atoms.
Solar reference abundances are from Savage and Sembach (1996), while B-star 
reference abundances are from Snow and Witt (1996),
Meyer et al. (1997a, 1997b), and Sofia et al. (1997).
$^{b}$  
The core composition is given by halo star abundances in Table 7 of Savage and
Sembach (1996).
$^{c}$  
From Savage and Sembach (1996).
\end{table}

%% file: tabhoppe.tex
\begin{table}
\begin{center}
\caption{ \label{hoppe.tab}
Types of Presolar Grains in Primitive Meteorites.}
\begin{tabular}{lcccc}
Mineral& Abund.& Size& Isotopic Signature& Stellar sources \\
&(ppm)& ($\mu$m)&& \\
\tableline
\\ 
Diamond& 1400& 0.002& Xe-HL& Type II supernovae \\
SiC Mainstream& 14& 0.1-20& enhancements$^{1}$ in $^{13}$C, $^{14}$N, & C-rich AGB stars \\
&&&$^{22}$Ne, heavy trace elements& \\
Graphite& 10& 0.8-12& enhancements in $^{12}$C, $^{18}$O& Type II supernovae, \\
&&&extinct $^{44}$Ti& (Wolf-Rayet stars) \\
Corundum& 0.3& 0.3-5& enhancements in $^{17}$O& Red giant, AGB stars \\
&&&depletion in $^{18}$O& \\

SiC X grains& 0.1& 0.5-10& enhancements in $^{12}$C, $^{15}$N, $^{28}$Si & Type II supernovae \\
&&&extinct  $^{26}$Al, $^{44}$Ti& \\

Silicon nitride& 0.002&$\sim$1& enhancements in $^{12}$C, $^{15}$N, $^{28}$Si & Type II supernovae \\
&&&extinct $^{26}$Al& \\
\end{tabular}
\end{center}
$^{1}$  Enhancement values are given relative to the Solar System isotopic composition. \\
\end{table}

%% file: table_ppm.tex
\begin{deluxetable}{clccccccc}
\footnotesize
\tablewidth{0pt}
\tablecaption{Elemental Abundances in Nearby Interstellar Gas \label{tab-ppm}}
\tablehead{
\colhead{}&\colhead{}&\colhead{}&\multicolumn{3}{c}{$<$ - - - - - -  Solar\tablenotemark{2}  - - - - - - $>$}
&\multicolumn{3}{c}{$<$ - - - - - -  B-star - - - - - - $>$}\nl
\colhead{Element}&\colhead{Log N}&\colhead{Denom.}&
\colhead{Ref.}&\colhead{Gas}&\colhead{Dust}&\colhead{Ref.}&\colhead{Gas}
&\colhead{Dust}\nl
\colhead{}&\colhead{cm$^{-2}$}&\colhead{PPM\tablenotemark{1,5}}&
\colhead{Abun.}&\colhead{PPM$_{\rm g}$\tablenotemark{1,10}}&
\colhead{PPM$_{\rm d}$\tablenotemark{1,6}}&
\colhead{Abun.}&\colhead{PPM$_{\rm g}$\tablenotemark{1,11}}&
\colhead{PPM$_{\rm d}$\tablenotemark{1,7}}
}

\startdata

{\bf $\epsilon$ CMa - LIC}\tablenotemark{3}&&&&&&&&\nl
C$^{+}$&14.20$^{+0.05}_{-0.06}$&H$^{\circ}$+H$^{\rm +}$&360&549$\pm83$&0&214&344$\pm66$&0\nl
N$^{\circ}$&13.20$\pm0.01$&H$^{\circ}$&93&79$\pm10$&14$\pm10$&75&50$\pm$8&25$\pm8$\nl
O$^{\circ}$&14.15$\pm0.03$&H$^{\circ}$&740&708$\pm100$&32$_{-32}^{+100}$&457&442$\pm73$&15$^{+73}_{-15}$\nl
Mg$^{\rm +}$&12.48$\pm0.01$&H$^{\circ}$+H$^{\rm +}$&38&10.5$\pm1.0$&28$\pm1$&25&6.5$\pm1.0$&18$\pm$1\nl
Si$^{\rm +}$+Si$^{\rm ++}$&12.87$\pm0.03$&H$^{\circ}$+H$^{\rm +}$&36&25$^{+6}_{-5}$&11$^{+6}_{-5}$&18.6&16$\pm4$&2.7$_{-2.7}^{+3.7}$\nl
S$^{\rm +}$&12.18--12.85&H$^{\circ}$+H$^{\rm +}$&19&5--25&0--14&12.3&3.3--15.5&0--9\nl
Fe$^{\rm +}$&12.13$\pm0.02$&H$^{\circ}$+H$^{\rm +}$&32&4.7$\pm0.4$&27$\pm0.4$&27&2.9$\pm0.4$&24.1$\pm0.4$\nl
&&&&&&&&\nl
{\bf$\lambda$ Sco}\tablenotemark{8}&&&&&&&&\nl
C$^{\rm +}$&15.70$^{+0.30}_{-0.70}$&H$^{\circ}$&360&295$^{+296}_{-239}$&65$^{+239}_{-65}$&214&295$^{+296}_{-239}$&0$^{+158}_{-0}$\nl
N$^{\circ}$\tablenotemark{11}&14.85$\pm$0.02&H$^{\circ}$&93&42$\pm5$&51$^{+5}_{-6}$&75&42$^{+6}_{-5}$&33$\pm5$\nl
O$^{\circ}$&15.86$^{+0.02}_{-0.01}$&H$^{\circ}$&740&427$^{+56}_{-53}$&313$^{+53}_{-56}$&457&427$^{+56}_{-53}$&30$^{+53}_{-30}$\nl
Si$^{\rm +}$&13.34$\pm$0.04&H$^{\circ}$&36&1.3$\pm$0.2&34.7$\pm0.2$&18.6&1.3$\pm0.2$&17.3$\pm$0.2\nl
S$^{\rm +}$&14.28$^{+0.06}_{-0.05}$&H$^{\circ}$&19&11$\pm2$&7.8$^{+1.8}_{-2.2}$&12.3&11$\pm2$&1.1$^{+1.8}_{-1.1}$\nl
Fe$^{\rm +}$&13.04$\pm$0.04&H$^{\circ}$&32&0.6$\pm$0.1&31.4$\pm0.1$&27&0.6$\pm0.1$&26.4$\pm$0.1\nl
&&&&&&&&\nl
\tablebreak
\tablenotetext{1}{PPM represents the number of atoms/ions per 10$^{6}$ 
hydrogen atoms.}
\tablenotetext{2}{The solar reference abundances 
are from Savage and Sembach (1996).
B-star abundances are from Meyer et al. (1997a, 1997b), Snow and Witt (1996),
and Sofia et al. (1997).}
\tablenotetext{3}{From --17 km s$^{-1}$ LIC component given in Gry et al. (1995), Gry and Dupin (1997), Dupin (1998) with uncertainties assumed to be $\pm$0.05 dex.}
\tablenotetext{4}{Estimated from $n$(e$^{\rm -}$)=$n$(p$^{\rm +}$)=0.1 cm$^{-3}$ and $n$(H$^{\circ}$)=0.22 cm$^{-3}$, giving log $N$(H$^{\circ}$+H$^{\rm +}$)=17.46 cm$^{-2}$.}
\tablenotetext{5}{This is the denominator used to calculate the gas-phase
PPM listed in columns 5 and 8 (see \S\ref{licgas2dust}).  Hydrogen column densities are from
Table \ref{tab-Rgd}.}
\tablenotetext{6}{The dust composition is derived from the assumption that the total elemental abundances in the cloud are given by solar abundances.} 
\tablenotetext{7}{The dust composition is derived from the assumption that the total elemental abundances in the cloud are given by B-star abundances.} 
\tablenotetext{8}{ From component 2 in York (1982).  This component is 
predominantly neutral and is coincident with the ``G'' velocity vector projected
in this direction if a correction of +5.6 km s$^{-1}$ is applied to the
{\it Copernicus} velocity scale.  This correction is required by 
Ca$^{\rm +}$ data, which find two clouds at --26$\pm$0.7 km s$^{-1}$ 
and --19.1$\pm$0.7 km s$^{-1}$ (Bertin et al.\ 1993).}
\tablenotetext{9}{$N({\rm H^{o}})$ is determined directly for $\lambda$ Sco (York 1983),
hence PPM$_{\rm g}$ does not depend on assumed reference abundance.}
\tablenotetext{10}{The denominator used to calculate this value is given in column 3 (with numerical values from Table 2).}
\tablenotetext{11}{The denominator used to calculate this value is given in column 3, with 
log $N$(H$^{\rm o}$)=17.505 cm$^{-2}$ and 
log $N$(H$^+$)=17.15 cm$^{-2}$.}
\enddata
\end{deluxetable}

%% file: table_star_abund.tex
\begin{deluxetable}{rccccccccl}
\small
\footnotesize
\tablewidth{0pt}
\tablecaption{Depletions in Directions of Nearby Stars \label{tab-femg}}
\tablehead{
\colhead{HR} 	& \colhead{Star}   & \colhead{Comp.}  & \colhead{V} &
\colhead{$N$(H$^{\circ}$)}  & \colhead{$N$(Fe$^{+}$)} & 
\colhead{$\delta _{\rm H}(Fe)$\tablenotemark{\*}} &
\colhead{$N$(Mg$^{+}$)}     & \colhead{$\delta _{\rm H}(Mg)$\tablenotemark{\*}}  &
\colhead{Refs.}\nl
\colhead{}&\colhead{}&\colhead{}&\colhead{km s$^{-1}$}&\colhead{cm$^{-2}$}&\colhead{cm$^{-2}$}&\colhead{dex}&\colhead{cm$^{-2}$}&\colhead{dex}&\colhead{}
}

\startdata
1708&$\alpha$ Aur&LIC&22&18.24$\pm$0.01&12.49$\pm$0.02&--1.26&12.85$\pm$0.02&--0.97&\tablenotemark{1,2}\nl
2491&$\alpha$ CMa&LIC&19.5&17.23$^{+0.17}_{-0.28}$&11.93$\pm$0.07&--0.98&12.20$\pm$0.08&--0.78&\tablenotemark{2,3,4}\nl
&&BC&13.7&17.23$\pm$0.17&11.73$\pm$0.09&--1.01&11.95$\pm$0.05&--0.86&\nl
2618&$\epsilon$ CMa&LIC&17&17.30\tablenotemark{11}&12.13$\pm$0.02&--0.72&12.48$\pm$0.014&--0.44&\tablenotemark{5}\nl
&&BC&10&16.88\tablenotemark{11}&11.72$\pm$0.04&--0.67&12.00$\pm$0.04&--0.46&\nl
2943&$\alpha$ CMi&LIC&20&17.88$\pm$0.01&12.05$\pm$0.02&--1.34&12.36$\pm$0.02&--1.1&\tablenotemark{1}\nl
&&No. 2&23&17.60$\pm$0.01&11.94$\pm$0.02&--1.17&12.11$\pm$0.03&--1.07&\nl
5191&$\eta$ UMa&&&17.85$\pm$0.17\tablenotemark{9}&12.43&--0.93&12.72&--0.71&\tablenotemark{6}\nl
5759&$\alpha$ Cen A&G&--18&18.0$\pm$0.02&12.45$\pm$0.02&--1.08&12.70$\pm$0.02&--0.88&\tablenotemark{2,7}\nl
5760&$\alpha$ Cen B&G&--18&18.0$\pm$0.02&&&12.72$\pm$0.08&--0.86&\tablenotemark{7}\nl
7557&$\alpha$ Aql&LIC&--17.4&18.26\tablenotemark{10}&12.57$\pm$0.12&&12.70$\pm$0.11&&\tablenotemark{2}\nl
&GD191-B2B&LIC&21&18.27&12.48$\pm$0.18&--1.30&12.72$\pm$0.14&--1.13&\tablenotemark{2,8}\nl
\tablenotetext{\*}{The depletions are calculated with respect to H$^{\rm o}$,
in this table, using solar abundances
of [Mg]=--4.42 dex and [Fe]=--4.49 dex.}
\tablenotetext{1}{Linsky et al. 1995}
\tablenotetext{2}{Lallement et al. 1995}
\tablenotetext{3}{Bertin et al. 1995}
\tablenotetext{4}{Lallement et al. 1994}
\tablenotetext{5}{Gry and Dupin 1997, Gry et al. 1995}
\tablenotetext{6}{Frisch 1998. The uncertainties on $N$(Fe$^{+}$) and $N$(Mg$^{+}$) are assumed values.}
\tablenotetext{7}{Linsky and Wood 1996}
\tablenotetext{8}{Lemoine et al. 1996}
\tablenotetext{9}{This value is determined from N$^{\circ}$ observations
and the assumption of solar abundances (\cite{fy91}).}
\tablenotetext{10}{Wayne Landsman, 1997, private communication.}
\tablenotetext{11}{For solar abundances; see Tables 2, 3.}
\enddata
\end{deluxetable}

%% file: bibliography.tex
%filename=bibliography.tex
%last update:  7/28/98